\documentstyle[12pt,epsfig]{article}

\advance\textheight by \topskip
\begin{document}
\tolerance=100000
\thispagestyle{empty}
\setcounter{page}{0}

\def\Dir{\kern -6.4pt\Big{/}}
\def\DDir{\kern -7.6pt\Big{/}}
\def\DGir{\kern -6.0pt\Big{/}}
\def\CCl{\kern -6.6pt^{(-)}}
%

\def\Ord{\buildrel{\scriptscriptstyle <}\over{\scriptscriptstyle\sim}}
\def\simlt{\rlap{\lower 3.5 pt \hbox{$\mathchar \sim$}} \raise 1pt \hbox {$<$}}
%

\def\OOrd{\buildrel{\scriptscriptstyle >}\over{\scriptscriptstyle\sim}}
\def\simgt{\rlap{\lower 3.5 pt \hbox{$\mathchar \sim$}} \raise 1pt \hbox {$>$}}
%

\def\sqr#1#2{{\vcenter{\hrule height.#2pt
      \hbox{\vrule width.#2pt height#1pt \kern#1pt
        \vrule width.#2pt}
      \hrule height.#2pt}}}
\def\square{\mathchoice\sqr68\sqr68\sqr{4.2}6\sqr{3.0}6}      
\def\leaderfill{\leaders\hbox to 1em{\hss.\hss}\hfill}
%

\font\fivesy=cmsy5 
\font\tensy=cmsy10 
%

\def\pscal{\hbox{\hskip2pt\fivesy\raise .5ex\hbox{\char'017}\hskip2pt }}
\def\pvet{\hbox{\hskip2pt\tensy\char'002\hskip -7.3pt\char'002\hskip2pt}}
%

\def\pmb#1{\setbox 0=\hbox{$#1$}%
\kern-0.25em\copy0\kern-\wd0
\kern0.5em\copy0\kern-\wd0
\kern-0.25em\raise.0433em\box0}
%

\def\epem{\ifmmode{e^+ e^-} \else{$e^+ e^-$} \fi}
\def\Im{\mathop{{\cal I}\mskip-4.5mu \lower.1ex \hbox{\it m}}}
\def\Re{\mathop{{\cal R}\mskip-4mu \lower.1ex \hbox{\it e}}}
%

\def\pl #1 #2 #3 {{\it Phys.~Lett.} {\bf#1} (#2) #3}
\def\np #1 #2 #3 {{\it Nucl.~Phys.} {\bf#1} (#2) #3}
\def\zp #1 #2 #3 {{\it Z.~Phys.} {\bf#1} (#2) #3}
\def\pr #1 #2 #3 {{\it Phys.~Rev.} {\bf#1} (#2) #3}
\def\prep #1 #2 #3 {{\it Phys.~Rep.} {\bf#1} (#2) #3}
\def\prl #1 #2 #3 {{\it Phys.~Rev.~Lett.} {\bf#1} (#2) #3}
\def\mpl #1 #2 #3 {{\it Mod.~Phys.~Lett.} {\bf#1} (#2) #3}
\def\rmp #1 #2 #3 {{\it Rev. Mod. Phys.} {\bf#1} (#2) #3}
\def\xx #1 #2 #3 {{\bf#1}, (#2) #3}
\def\preprint{{\it preprint}}
\def\sm{${\cal {SM}}$}
\def\mssm{${\cal {MSSM}}$}
%
\def\etal {{\it et al.}}
\def\eg {{\it e.g.}}
\def\ie {{\it i.e.}}
\def\ibid{{\it ibid.}}
%

\def\roots{\mbox{$\sqrt{s}$}}
\def\MSbar{\mbox{$\overline{MS}$}}
\def\ccbar{\mbox{$c \bar{c} $}}
\def\bbbar{\mbox{$b \bar{b} $}}
\def\ttbar{\mbox{$t \bar{t} $}}
\def\Y{\mbox{$\Upsilon$}}
\def\aem{\mbox{$\alpha_{{\rm em}}$}}
\def\shat{\mbox{$\hat{s}$}}
\def\lhat{\mbox{$\hat{\cal L}$}}
\def\sighat{\mbox{$\hat{\sigma}$}}
\def\rs{\mbox{$\sqrt{s}$}}
\def\pT{\mbox{$p_T$}}
\def\mjj{\mbox{$M_{jj}$}}
\def\wgg{\mbox{$W_{\gamma \gamma}$}}
\def\fge{\mbox{$f_{\gamma|e}$}}
\def\fee{\mbox{$f_{e|e}$}}
\def\ptmin{\mbox{$p_{T,min}$}}
\def\qvec{\mbox{$\vec{q}^{\gamma}$}}
\def\xq{\mbox{$(x,Q^2)$}}
\def\qig{\mbox{$q_i^{\gamma}$}}
\def\Gg{\mbox{$G^{\gamma}$}}
\def\qqbar{\mbox{$q \bar{q}$}}
\def\ppbar{\mbox{$p \bar{p}$}}
\def\gaga{\mbox{$\gamma\gamma$}}
\def\be{\begin{equation}}
\def\ene{\end{equation}}
\def\ba{\begin{eqnarray}}
\def\ena{\end{eqnarray}}
\def\een{\end{subequations}}
\def\ben{\begin{subequations}}
\def\beq{\begin{eqalignno}}
\def\eeq{\end{eqalignno}}
\def\ee{e^+e^-}
\def\to{\rightarrow}
\def\bbar{{\bar{b}}}
\def\bb{{b\bar{b}}}
\def\tbar{{\bar{t}}}
\def\degree{^{\circ}}
\def\fm{{\rm fm}}
\def\MeV{{\mbox{MeV}}}
\def\GeV{{\mbox{GeV}}}
\def\TeV{{\mbox{TeV}}}
\def\cM{{\cal M}}
\def\cR{{\cal R}}
\def\as{\alpha_s}
\def\gt{\Gamma}
\def\eps{\varepsilon}
\def\non{\nonumber}
\def\Li{\mbox{\rm Li}}
\def\chitil{\widetilde\chi}
\def\tauptaum{\tau^+\tau^-}
\def\mhp{m_{\hp}}
\def\hp{H^+}
\def\hm{H^-}
\def\hpm{H^{\pm}}
\def\mhpm{m_{\hpm}}
\def\zsm{Z}
\def\mzsm{m_{\zsm}}
\def\wpm{W^{\pm}}
\def\mwpm{m_{\wpm}}
\def\mt{m_t}
\def\mb{m_b}
\def\mc{m_c}
\def\hl{h^0}
\def\hh{H^0}
\def\ha{A^0}
\def\mhl{m_{\hl}}
\def\mhh{m_{\hh}}
\def\mha{m_{\ha}}
\def\eps{\epsilon}
\def\sutwoone{SU(2)\times U(1)}
\def\sutwo{SU(2)}
\def\lam{\lambda}
\def\munichnlc{{\it Proceedings of the Workshop on ``$\epem$ Linear Colliders
at $500\gev$: the Physics Potential''},  publication DESY 92-123A (1992),
ed. P.M. Zerwas,
Feb. 4 - Sept. 3 (1991) -- Munich, Annecy, Hamburg}
\def\saariselka{{\it Proceedings of the 1st International Workshop
on ``Physics and Experiments with Linear $\epem$ Colliders''},
eds. R. Orava, P. Eerola, and M. Nordberg, Saariselka, Finland,
September 9-14, 1992 (World Scientific Publishing, Singapore, 1992)}
\def\hawaii{{\it Proceedings of the 2nd International Workshop on
``Physics and Experiments with Linear $\epem$ Colliders''}, ed. F. Harris,
Waikoloa, HI, April 26-30, 1993}
\def\doeack{\foot{Work supported, in part, by the Department of Energy.}}
\def\gam{\gamma}
\def\Gam{\Gamma}
\def\lamnew{\Lambda}
\def\QS{{Q^2}}
\def\cale{{\cal E}}
\def\calo{{\cal O}}
\def\calm{{\cal M}}
\def\cala{{\cal A}}
\def\stop{{\wtilde t}}
\def\mstop{m_{\stop}}
\def\ep{e^+}
\def\em{e^-}
\def\ptr{p_T}
\def\mmissl{M_{miss-\ell}}
\def\mmisslmin{\mmissl^{\rm min}}
\def\gev{~{\rm GeV}}
\def\tev{~{\rm TeV}}
\def\pbi{~{\rm pb}^{-1}}
\def\fbi{~{\rm fb}^{-1}}
\def\fb{~{\rm fb}}
\def\pb{~{\rm pb}}
\def\smv{{\it Proceedings of the 1990 DPF Summer Study on
High Energy Physics: ``Research Directions for the Decade''},
editor E. Berger, Snowmass (1990)}
\def\perspectives{{\it Perspectives on Higgs Physics}, ed. G. Kane, World
Scientific Publishing, Singapore (1992)}
\def\ibid{{\it ibid.}}
\def\mt{m_t}
\def\wp{W^+}
\def\wm{W^-}
\def\rta{\rightarrow}
\def\tanb{\tan\beta}
\def\sinb{\sin\beta}
\def\cosb{\cos\beta}
\def\cotb{\cot\beta}
\def\lplm{l^+l^-}
\def\cale{{\cal E}}
\def\calo{{\cal O}}
\def\calm{{\cal M}}
\def\mpp{\calm_{++}}
\def\mmm{\calm_{--}}
\def\mpm{\calm_{+-}}
\def\mmp{\calm_{-+}}
\def\absq#1{\left| #1 \right|^2}
\def\im{{\rm Im}\,}
\def\re{{\rm Re}\,}
\def\hn{h}
\def\hsm{\phi^0}
\def\mhn{m_\hn}
\def\mhsm{m_{\hsm}}
\def\nsd{N_{SD}}
\def\dg{\Delta g}
\def\du{\Delta u}
\def\dd{\Delta d}
\def\ds{\Delta s}
\def\dgam{\Delta \Gamma}
\def\mw{m_W}
\def\mz{m_Z}
\def\anti{\bar}
\def\fhalfs#1{F^s_{1/2}(#1)}
\def\fhalfp#1{F^p_{1/2}(#1)}
\def\ifmath#1{\relax\ifmmode #1\else $#1$\fi}
\def\half{\ifmath{{\textstyle{1 \over 2}}}}
\def\sixth{\ifmath{{\textstyle{1 \over 6}}}}
\def\threeeighths{\ifmath{{\textstyle{3 \over 8}}}}
\def\fivehalfs{\ifmath{{\textstyle{5 \over 2}}}}
\def\quarter{\ifmath{{\textstyle{1 \over 4}}}}
\def\onenineth{\ifmath{{\textstyle{1\over 9}}}}
\def\fournineths{\ifmath{{\textstyle{4\over 9}}}}
\def\3quarter{{\textstyle{3 \over 4}}}
\def\third{\ifmath{{\textstyle{1 \over 3}}}}
\def\twothirds{{\textstyle{2 \over 3}}}
\def\fourthirds{{\textstyle{4 \over 3}}}
\def\eightthirds{{\textstyle{8 \over 3}}}
\def\fourth{\ifmath{{\textstyle{1\over 4}}}}
\def\nicefrac#1#2{\hbox{${#1\over #2}$}}
\def\eps{\epsilon}
\def\ebtag{e_{b-tag}}
\def\emistag{e_{mis-tag}}
\def\tth{t\anti t \hn}
\def\ttz{t\anti t Z}
\def\ttglnu{t\anti t g-\ell\nu}
\def\ttbdecay{t\anti t g-b\ell\nu}
\def\ar{\rightarrow}
\def\F{\ifmmode{\cal F}\else{$\cal F$}\fi}
\def\X{\ifmmode{\cal X}\else{$\cal X$}\fi}
\def\Y{\ifmmode{\cal Y}\else{$\cal Y$}\fi}
\def\Z{\ifmmode{\cal Z}\else{$\cal Z$}\fi}
\def\li{\ifmmode{p_i,\lambda}\else{$p_i,\lambda$}\fi}
\def\lj{\ifmmode{p_j,\lambda'}\else{$p_j,\lambda'$}\fi}
\def\l #1{\ifmmode{p_{#1},\lambda_{#1}}\else{$p_{#1},\lambda_{#1}$}\fi}
\def\m #1{\ifmmode{q_{#1},\lambda_{#1}}\else{$q_{#1},\lambda_{#1}$}\fi}
\def\r #1{\ifmmode{r_{#1},-}\else{$r_{#1},-$}\fi}
\def\MH{\ifmmode{M_H}\else{$M_H$}\fi}
\def\MZ{\ifmmode{M_Z}\else{$M_Z$}\fi}
\def\MW{\ifmmode{M_W}\else{$M_W$}\fi}
%
\def\ttH{\ifmmode{t\bar tH}\else{$t\bar tH$}\fi}
\def\WH{\ifmmode{W H}\else{$W H$}\fi}
\def\ZH{\ifmmode{ZH}\else{$ZH$}\fi}
\def\ggf{\ifmmode{gg\ar H}\else{$gg\ar H$}\fi}
\def\WWH{\ifmmode{W^+W^-\ar H}\else{$W^+W^-\ar H$}\fi}
\def\ZZH{\ifmmode{ZZ\ar H}\else{$ZZ\ar H$}\fi}
\def\ph{\ifmmode{\gamma}\else{$\gamma$}\fi}
\def\Hphph{\ifmmode{H\rightarrow\gamma\gamma}
           \else{$H\rightarrow\gamma\gamma$}\fi}
\def\Hbb{\ifmmode{H\rightarrow b\bar b}
           \else{$H\rightarrow b\bar b$}\fi}
\def\Hfourl{\ifmmode{H\rightarrow Z^{*}Z^{*}\rightarrow  4\ell}
            \else{$H\rightarrow Z^{*}Z^{*}\rightarrow 4\ell$}\fi}
\def\Hllnn{
\ifmmode{H\rightarrow ZZ\rightarrow \ell^+\ell^-\nu_{\ell'}\bar\nu_{\ell'}}
\else{$H\rightarrow ZZ\rightarrow \ell^+\ell^-\nu_{\ell'}\bar\nu_{\ell'}$}\fi}
\def\Hlljj{
  \ifmmode{H\rightarrow ZZ\rightarrow \ell^+\ell^-{{jj}}}
  \else{$H\rightarrow ZZ\rightarrow \ell^+\ell^-{{jj}}$}\fi}
\def\Hlnjj{
  \ifmmode{H\rightarrow WW\rightarrow \ell\nu_\ell~jj}
  \else{$H\rightarrow WW\rightarrow \ell\nu_\ell~jj$}\fi}
\def\Hphpha{{1a}}
\def\Hphphb{{2a}}
\def\Hbba{{3a}}
\def\Hbbb{{4a}}
\def\Hfourla{{5}}
\def\Hfourlb{{6}}
\def\Hfourma{{7}}
\def\Hfourmb{{8}}
\def\Hfourlaheavy{{9}}
\def\Hfourlbheavy{{10}}
\def\Hfourlaobese{{11}}
\def\Hfourlbobese{{12}}
\newcommand{\nn}{\noindent}
\def\mathrm{\rm}

\begin{flushright}
{\large Cavendish--HEP--95/18}\\ 
{\large DFTT 79/95}\\ 
{\large DTP/95/104}\\ 
{\rm November 1996\hspace*{.5 truecm}}\\ 
\end{flushright}

\vspace*{\fill}

\begin{center}
{\Large \bf Higgs Production at the LHC:\\
an Updated Signal-to-Background Analysis}\\[2.cm]
{\large Stefano Moretti\footnote{E-mails: Moretti@hep.phy.cam.ac.uk.;
Moretti@to.infn.it}}\\[0.5 cm]
{\it Cavendish Laboratory, 
University of Cambridge,}\\ 
{\it Madingley Road,
Cambridge, CB3 0HE, United Kingdom.}\\[0.5cm]
{\it Dipartimento di Fisica Teorica, Universit\`a di Torino,}\\
{\it and I.N.F.N., Sezione di Torino,}\\
{\it Via Pietro Giuria 1, 10125 Torino, Italy.}\\[0.75cm]
\end{center}

\vspace*{\fill}

\begin{abstract}
{\normalsize
\noindent
This paper follows Ref.~\cite{signals}, where updated calculations of 
cross sections and branching ratios relevant for Standard Model Higgs 
phenomenology at the LHC were presented. Here, we complete that study 
by carrying out an updated signal-to-background analysis. We present
results obtained by using
exact matrix element computations 
at parton level for all processes, by exploiting 
the most recent parton distributions fitted to HERA structure function
data and the most recent values of 
the electroweak input parameters. Cross sections
and distributions are given for two collider energies, $\sqrt s_{pp}=10$ TeV and
14 TeV. Event rates and significances are discussed for two possible values
of integrated luminosity, 10 fb$^{-1}$ and 100 fb$^{-1}$.}
\end{abstract}

\vspace*{\fill}
\newpage
\section{Introduction}

The search for the Standard Model (\sm) 
Higgs boson after LEP~2 and before a Next Linear Collider
(NLC) relies on the Large Hadron Collider (LHC). 
The project of a proton-proton collider at CERN dates back to the
end of the $Sp(\bar p)S$ era, 
when the need of a hadron accelerator operating at the TeV scale
in order to carefully study the Higgs sector of the \sm\ clearly
came out. 
At the beginning, the LHC was planned as a 
machine with a beam energy of 7.7 TeV and a design
luminosity of $1.7\times10^{34}~{\mbox{cm}}^{-2}~{\mbox{s}}^{-1}$
or more \cite{DATA}.  

Since the time of the 1990 Large Hadron Collider Workshop \cite{LHC},
when  all the most important physics issues which can be studied at
the CERN $pp$ accelerator were addressed in great detail
(Higgs physics included), both the design of the machine and its 
foreseen performances
have partially evolved. For example, the Centre-of-Mass (CM) energy has been  
reduced. In fact, the CERN
Council finally decided in December 1994 that the LHC should be
built as a two-stage project. The first stage being a particle collider
with a CM energy of 10 TeV, which should be ready to start running in
2004. In 2008 the LHC should be upgraded by adding magnets to reach
the final value $\sqrt s_{pp}=14$ TeV. Not before 1997 it could be re-examined
the two stage project, exploiting the possibility of reverting to the
immediate construction of a 14 TeV accelerator to be ready by 2005.
This however depends on the availability of sufficient financial 
commitments. 
In addition, the peak design luminosity has decreased as well, to
$1.0\times10^{34}~{\mbox{cm}}^{-2}~{\mbox{s}}^{-1}$ or a factor of ten less.

In the meanwhile, also the evolution of the LHC detectors has been greatly 
carried out, as new technologies have become available in the last six years. 
In particular, for what concerns Higgs
physics, special attention 
has been devoted to the combined performances of: (i) the 
electromagnetic
calorimeter and the inner detector, in view of higher mass resolutions
in tagging $H\ar ZZ^*\ar 4e$ and $H\ar \gamma\gamma$ signals,
of more efficient $\gamma/\pi^0$ and $\gamma/$jet separation, and
better  $b$-tagging, especially using low $p_T$ electrons;
(ii) the muon chamber and the inner detector for studies of $H\ar ZZ^*\ar 4\mu$
with high mass resolution. 

Recently, the Higgs discovery potential of the LHC has
been carefully re-investigated in the Technical Proposals 
of the two experimental collaborations
that will work at the CERN hadron collider: ATLAS \cite{ATLAS}
and CMS \cite{CMS}. There, all the relevant information concerning
the design/performances
of the machine and of the detectors were also given.
However, concerning the physics aspects of those analyses,
a few important features must be noticed: first,
predictions were given for the CM energy of 14 TeV only; furthermore,
one often finds there slight
inconsistencies in the way the signal and background rates were
calculated and compared with each other
(leading versus next-to-leading order cross sections,
out-of-date parton distributions and parameter values, etc ...).
Moreover, sometimes event rates were computed using exact matrix
element calculations, whereas in other instances Monte Carlo event generators
were used (such as, e.g.,
PYTHIA, HERWIG, GEANT \cite{generators}).

For this reasons, in Ref.~\cite{signals}, cross sections, branching ratios
and event rates of signals relevant for Higgs phenomenology at the LHC
were all recomputed. In that paper, all the most recent theoretical 
results and experimental inputs were included, such as:
\begin{itemize}
\item[{(i)}] next-to-leading (NLO) order corrections to most of the 
Higgs production cross sections and partial decay widths;
\item[{(ii)}] new parton distribution functions, fitted to 
the high precision data mostly coming from deep inelastic scattering at HERA
and to the latest measurements of $\alpha_s$;
\item[{(iii)}] new input parameter values 
of physical quantities (in particular the top quark mass $m_t$), obtained
from precision
measurements performed at LEP, Tevatron and other
machines.
\end{itemize}
The goal of the analysis carried out there was a set of benchmark results 
for cross sections and event rates as a function of $M_H$, for the two
`standard' LHC collision energies, $\sqrt{s} = 10$ and $14~{\mbox{TeV}}$.
The predictions given in Ref.~\cite{signals} 
enable the Higgs production and background 
rates as well as the significance factors
used, e.g.,  in Refs.~\cite{ATLAS} and \cite{CMS}, to be normalised to the
most up-to-date values.

This paper is, in a sense, the continuation of Ref.~\cite{signals}, as it
contains an updated signal-to-noise analysis, which takes
into account most of the theoretical improvements adopted in 
Ref.~\cite{signals}
as well as the new characteristics of the collider discussed above. 
However, as already mentioned in Ref.~\cite{signals}, we again stress
here that we are not attempting to perform a detailed analysis of
signals, backgrounds and search strategies: in this respect, by far the
most complete studies to date can still be found in the ATLAS
\cite{ATLAS} and CMS \cite{CMS} Technical Proposals. 
Instead, it is the purpose of the
present study to approach the issue of phenomenological
studies of Higgs signals and backgrounds
at the LHC in the most coherent way,
preferring {\sl incompleteness} to {\sl inconsistency}.
In fact, throughout this paper, only {\sl 
exact matrix element computations}
are employed for {\sl all processes}, 
both signal and background events, and no recourse to any
Monte Carlo event generator is done.  Not even fragmentation and
hadronization phenomena are considered and no special effort 
in simulating detector effects (minimum bias events,
smearing, pile-up, etc ...) is done either.
All results are computed and cuts are applied at {\sl parton level}. 
In evaluating the Feynman amplitude 
squared of the various processes, if not otherwise stated, we used the
packages MadGraph \cite{MadGraph} and HELAS \cite{helas} and the integrator
VEGAS \cite{Vegas}.

There is however a difference, with respect to Ref.~\cite{signals}.
As the most part of the backgrounds studied in literature 
have not benefited so far
from much theoretical effort (contrary to the Higgs processes), such that
most of them are known at tree--level only,
we have adopted also for the signal rates the 
leading order (LO) results (contrary to Ref.~\cite{signals}). 
Indeed, in the spirit of the approach we described, we have `independently' 
computed here all background processes using perturbative techniques
implementing standard Feynman rules.

Concerning the numerical part of our work, we have used 
for the electroweak and QCD  input parameters the same values
given in Ref.~\cite{signals}. For reference we list
them here too, they are:
$$M_Z=91.186~{\mathrm {GeV}},\quad\quad \Gamma_Z=2.495~{\mathrm {GeV}},$$
$$M_W=80.356~{\mathrm {GeV}},\quad\quad \Gamma_W=2.088~{\mathrm {GeV}},$$
\be\label{ewparam}
G_F=1.16639\times10^{-5}~{\mathrm {GeV}}^{-2},
\quad\quad\alpha_{em}\equiv \alpha_{em}(M_Z)= 1/128.9.
\ene
The charged and neutral weak fermion--boson couplings are defined by
\be\label{GF}
g_W^2 = \frac{e^2}{\sin^2\theta_W} =  4 \sqrt{2} G_F M_W^{2}, \qquad
g_Z^2 = \frac{e^2}{\sin^2\theta_W\; \cos^2\theta_W} =  4 \sqrt{2} G_F M_Z^{2}.
\ene
For the vector and axial couplings
of the $Z$ boson to fermions, we use the `effective leptonic' value
$\sin^2_{\mathrm {eff}}(\theta_W)=0.2320$.
For the QCD strong coupling constant $\alpha_s$
we always adopt the expression
at two-loops, with $\Lambda^{(4)}_{\overline{{MS}}}=230$~MeV, in order
to match  our default parton distribution set MRS(A) \cite{MRSA},
and with a scale $\mu$ set equal to the
the subprocess invariant mass,
$\mu = \sqrt{\mathaccent 94{s}}_{pp}$.
For the fermion masses we  take
$m_\mu=0.105$~GeV, $m_\tau=1.78$~GeV, $m_s=0.3$~GeV, $m_c = 1.4$~GeV,
 $m_b=4.25$~GeV
and $m_t=175$ GeV \cite{CDFtop,D0top}, 
with all decay widths equal to zero except
for $\Gamma_t$. We calculate this at tree-level
within the \sm, using the expressions
given in Ref.~\cite{widthtopSM}.
The first generation of fermions and all neutrinos are taken to be
massless, i.e. $m_u=m_d=m_e=m_{\nu_e}=0$ and
$m_{\nu_\mu}=m_{\nu_\tau}=0$. Finally, we consider Higgs masses spanning
in the range $80\ {\mathrm{GeV}} \Ord M_H \Ord 700$ GeV, assuming that 
the first value is the (conservative) discovery limit for 
LEP2\footnote{Note that the current lower limit on the Higgs mass from
direct searches is 66 GeV \cite{lower}, whereas from the  
fits to the LEP and SLD data one can deduce a 95\% confidence level
upper limit on $M_H$ of 550 GeV (with the best $\chi^2$ fit for 
$M_H=149^{+148}_{-82}$ GeV) \cite{blondel}.}. As usual, our 
discussion will be subdivided into two distinct classes, depending on whether
$M_H$ is less than (i.e., `intermediate mass' range) or greater than 
(i.e., `heavy mass' range) the $WW$-decay threshold around $2 M_W$.

The plan of this paper is as follows. In Section 2 
we review the most important Higgs signatures (see Ref.~\cite{signals})
for a \sm\ Higgs in the intermediate mass range. That is, 
we will  proceed by studying the channels:
\begin{itemize}
\item \Hphph;
\item \Hbb;
\item $H\ar ZZ^*\ar 4\ell$ ($\ell=e,\mu$);
\end{itemize}
and the corresponding 
backgrounds\footnote{Very recently 
(see Ref.~\cite{dittdrei}) it has been claimed
that the decay
$H\ar W^{(*)}W^{(*)}\ar \ell^+\nu_\ell \ell'^-\bar\nu_{\ell'}$, where
$\ell,\ell'=e,\mu$, can give additional chances of Higgs detection in
the window $155~{\mathrm{GeV}} \Ord M_H  \Ord 180~{\mathrm{GeV}} $.  
In this case the lack of a measurable narrow resonant
peak should be compensated by a relatively large branching ratio,
since for the above mass range the $WW$-channel is the dominant decay mode.}.
In Section 3 we concentrate
on the heavy mass range, in particular, 
we will  consider the following channels:
\begin{itemize}
\item $H\ar ZZ\ar 4\ell$ ($\ell=e,\mu$);
\item \Hllnn\ ($\ell=e,\mu$ and $\ell'=e,\mu,\tau$);
\end{itemize}
and again the corresponding backgrounds.
Finally, in Section 4 we give a short summary 
and draw our conclusions.

\section{The intermediate mass range}

\subsection{Search for $H\ar \gam\gam$}
\label{sec:phph}

The importance of the rare decay mode 
$H\ar \gamma\gamma$
has been clearly explained in Ref.~\cite{signals}. The highest rates for
this channel occur in the region 80--150 GeV, where the combination
of a rising branching ratio (see Fig.~1 in Ref.~\cite{signals}) and  a 
falling cross-section (see Figs.~5a--5b in Ref.~\cite{signals}) 
yields rates  which are remarkably
constant over the above range. As it has been recently outlined
that the region 80--100 GeV could be better probed by the channel
$\Hbb$, whereas for $\MH\OOrd130$ GeV the channel $\Hfourl$ becomes 
clearly visible, the two-photon decay should remain the best way to
search for the Higgs boson in the mass interval 100--130 GeV \cite{last}.

It has been shown by ATLAS \cite{ATLAS}
that the potential interest of the inclusive
channel \Hphph\ (without isolated and high $p_T$ leptons) would
be limited to the region 80--100 GeV, and only after several years of data
taking at high luminosity. However,
since large part of this interval can be covered by LEP~2 (with
$\sqrt s_{ee}\approx200$ GeV) and/or Di--Tevatron (with 
$\sqrt s_{pp}\approx4$ TeV), and  
since the bulk of the events at high significance would come anyway from 
the signature $\ell\gam\gam X$ (with an isolated high $p_T$
lepton), we do not consider here
this case. Somehow more optimistic prospects are given by CMS
\cite{CMS}, which also considered the channel \Hphph\ in association
with high $E_T/E$ jets. 

We concentrate here on the associated 
production of the $H$ with a $t\bar t$-pair or a $W$,
followed by the (semi)leptonic decays $t\bar t\ar \ell\bar\nu_\ell X$ 
and $W\ar \ell\bar\nu_\ell$, respectively. 
These channels have substantially different 
characteristics with respect to the case of the direct $gg\ar\Hphph$
production and decay. On the one hand, one has the disadvantage
of production rates which are one order of magnitude smaller and 
a larger number of backgrounds (both reducible and irreducible).
On the other hand, the primary vertex
position can be more easily worked out, by using the charged lepton track 
\cite{ATL17} and, in addition, the isolated hard lepton from the $W$ and $t$
decays allows for a very strong reduction of the backgrounds. Finally, 
these advantages can be exploited both at high and low luminosity.  

In order to account for the main features of the 
$WH\ar\ell\gam\gam X$ and $\ttH\ar
\ell\gam\gam X$ signals, detectors
must guarantee the following performances \cite{CMS}:
\begin{itemize}
\item an electromagnetic calorimeter with excellent stability,
uniformity and high energy and angular
resolution, in order to extract the very narrow Higgs resonance
in two photons  from the continuum $\gam\gam$ background;
\item a large acceptance;
\item excellent neutral pions (giving photons) rejection, since QCD
jets contain one or more leading $\pi^0$'s (it has been estimated that
a rejection factor $\OOrd 5\times10^3-10^4$ should allow one to reduce
the QCD backgrounds below the $\gam\gam$ continuum);
\item powerful capability of the inner tracking system. 
\end{itemize} 
A very detailed description of the potential of the LHC detectors
needed in order to make feasible to tag the Higgs boson in the
di-photon channel can be found in  
Ref.~\cite{gamgam}. Although
this study dates back at the time of the first LHC Workshop (1990),
nevertheless it still represents an excellent source of informations on
this topic. We certainly refer the reader to that paper.
In practice, what it is especially needed is that the LHC detectors
can maintain an excellent photon energy and angular resolution, while
the machine is running at full luminosity. 

After applying standard acceptance cuts (see below),
the main backgrounds to the $\ell\gam\gam$ Higgs signature via $WH$
and $\ttH$ production have found to be  the irreducible processes $W\gam\gam$ 
and $t\bar t\gam\gam$\footnote{In generating the results for
the background $t\bar t\gam\gam$ we have used the code
already employed in Ref.~\cite{GattoeVolpe}, for the subprocess
$gg\ar t\bar t \gam\gam$, whereas for the case  $t\bar t\gam$ it has
been employed the program used in Ref.~\cite{primo} (for both $gg$-
and $q\bar q$-fusion).} 
and the reducible reactions $W\gam j$ 
with the pions in the jet giving hard photons, 
$t\bar t\gam$ with a jet and
$Z\gam$ with a $\ell^\pm$ from the $Z$-decay faking a photon, respectively.
It has been shown in Ref.~\cite{CMS} 
that the
processes $Wjj$ and $WZ$ as well as the
channels $\ttbar\gam j$, $\bbbar\gam\gam$, $\ccbar\gam\gam$, $\bbbar\gam j$
and $\ccbar\gam j$ give a minor contribution, so we dot treat them here. 

The isolation criteria and cuts we have applied to select the signals are
the same ones adopted by the CMS Collaboration in its Technical
Proposal for the LHC \cite{CMS}. We list them here for convenience:

\begin{itemize}
\item for photons, $|\eta^\gam|<2.4$, $p_T^{\gamma_1}>40~\GeV$, 
$p_T^{\gamma_2}>20~\GeV$;
\item for leptons, $|\eta^\ell|<2.4$, $p_T^{\ell}>20~\GeV$;
\item isolation of leptons and photons, i.e., no particles with $p_T>2$ GeV
in a azimuthal angle-(pseudo)rapidity cone of $\Delta R=
\sqrt{\Delta\eta^2+\Delta\phi^2}$,
\end{itemize}
where the subscripts 1 and 2 for the photons refer to the most and
least energetic one, respectively. 
In the case of the background $Z\ph$ the additional
requirement $|M_{Z}-M_{\ell\gam}|>5$ GeV was also applied.

Figs.~\Hphpha--\Hphphb\ show the differential distribution of the
sum of the signals and of the backgrounds, respectively, in the invariant mass
$M_{\gam\gam}$,
for $\sqrt s_{pp}=10, 14$ TeV and $m_t=175$ GeV.
We have collected the signal rates in bins of 2 GeV around the Higgs mass
(shown for the values 80, 100 and 120 GeV), as the Higgs
resonance in this
region is very narrow (e.g., $\Gamma_H\Ord0.01~\GeV$).
In the plots all the above cuts have been applied.
They also include the suppression factor against $\pi^0$'s in jets faking
photons\footnote{Which is equal to 
$\approx5000$ for $W\gamma$j and $t\bar t\gam$ events, see 
Ref.~\cite{gamgam} and also Ref.~\cite{ATLAS}.} and the fact that
only 1\% of the leptons from the $Z$ decay in $WZ\ar
W\ell^+\ell^-$ are recognised as photons \cite{ATLAS,CMS}
(the factor $\varepsilon$ in the figures gives account of this).
We have not considered the conversion loss, of
$\calo(10\%-20\%)$ approximately, for tagging real photons \cite{ATLAS,CMS}.

In Tab.~I we have
collected the number of signal $S$ and background $B$
events in a window of $\Delta M\equiv|M_H-M_{\gam\gam}|=1$ GeV 
around the actual values
of $\MH$, for the integrated luminosity 
${\cal{L}}=100$~fb$^{-1}$. Significances $S/\sqrt B$ are given for
$\Delta M$ equal to 1, 2 and 3 GeV.
 
At $\sqrt s_{pp}=10$ TeV, values of $S/\sqrt B$ are such that in the
case the high luminosity option and good resolution in
$M_{\gam\gam}$ can be contemporaneously guaranteed, i.e., ${\cal
L}=100$ fb$^{-1}$ and $\Delta M\Ord 3$ GeV, then Higgs detection
should be promptly feasible for values of $\MH$ in the range 80--120 GeV.
Clearly, if the high luminosity option and/or high mass resolutions
will not be possible, for $\sqrt s_{pp}=10 $ TeV, such chances would
sensibly decrease (for example, for  ${\cal{L}}=10$~fb$^{-1}$, the
significances in Tab.~I are reduced by a factor ${\sqrt{10}}\approx3.2$).  
This is particularly true for a light Higgs.

At $\sqrt s_{pp}=14 $ TeV values of  $S/\sqrt B$ are larger by a
factor of 1.3--1.4, if compared to those at 10 TeV, independently
of $\MH$ and $m_t$. This guarantees further possibilities of Higgs
discovery, since even for very modest resolutions in $M_{\gam\gam}$ 
(although CMS claims that the mass 
resolution achievable could vary between
0.8 and 1.1 GeV \cite{CMS} !) and/or a reduced luminosity, only the case 
 $\MH\approx 80$ GeV would present some difficulties. 

The relative contribution of the various background sources (before
including any efficiency $\varepsilon$) can be appreciated in Figs.~1b and 2b
(for $\sqrt s_{pp}=10$ and 14, respectively). 

\subsection{Search for $H\ar b\bar b$}
\label{sec:bb}

An alternative strategy in searching for the \sm\ Higgs boson in the
intermediate mass range is to look for its main decay channel $\Hbb$,
by resorting to techniques of $b$-flavour identification \cite{btagg,SDC},
thus reducing the enormous QCD backgrounds of light quark and gluon jets.
This channel is the dominant decay mode in the range 80 GeV
$\Ord\MH\Ord$ 130 GeV (see Fig.~1 of Ref.~\cite{signals}). 
Around $\MH=130$ GeV the decay $H\ar
WW^{*}$ starts dominating. The possibility
of selecting the $b\bar b$ channel out of the huge QCD background
by using the $b$--tagging capabilities of
vertex detectors  was first suggested in some
theoretical papers \cite{Tevadetect1,Tevadetect2}. 
The main difficulties in this kind of search are the
 expected low Higgs rates after signal reconstruction, the necessity of 
gaining 
an accurate control of all the numerous background sources and the one of
achieving very high $b$--tagging performances \cite{last}.\par
The chances to tag the \sm\ Higgs boson via its main decay channel in
the intermediate mass range are provided by the associate production
mechanisms $\WH$ and $\ttH$, with the $W$ and one of the top quarks
decaying semileptonically to electrons or muons. The lepton is usually
at high $p_T$ and isolated, such that it can be used for triggering
purposes. The expected signatures
would then be $\ell b\bar bX$ (from $WH$) and $\ell b \bar bb\bar bX$ 
(from $\ttH$).
Higgs signals in the $\bbbar$ channel
would appear as a (narrow) peak in the invariant mass
distribution of $b$-quark pairs.

In addition to the associated production with a charged vector boson
and a $\ttbar$ pair, also other \sm\ production mechanisms and signatures
involving the decay $\Hbb$
have been suggested in literature. We list them below for completeness
although we will not treat them here, for the reasons that we are going to
illustrate.
\begin{itemize}
\item $\ZH$ production \cite{gny} followed by the decays $Z\ar \ell^+\ell^-$ 
($\ell=e,\mu$) and $\Hbb$. The disadvantages of this case are that, on
the one hand, the production cross section is approximately six times
lower that in the case of $\WH$ production whereas, on the other hand,
the irreducible background $Z\bbbar $ is only $\approx1.8$ times
smaller than the corresponding $W\bbbar$ background to $\WH$ production. 
\item  $\ZH$ production \cite{gny} followed by the decays $Z\ar
\nu\bar\nu$ and  $\Hbb$, as suggested in Ref.~\cite{Tevadetect1} (in the
second paper).
In this case it is very hard matter to reconstruct the final state 
\cite{ATLAS}, because of the presence of two neutrinos escaping
the detectors. In addition, background processes with
$E_T^{{miss}}$ are large, if compared to the (rather low) signal.
\end{itemize}

Like in the case of the decay $\Hphph$, the challenge of the LHC
in selecting the signature $\Hbb$ requires very high
performances from the detectors. 
Here we list the set of cuts we have adopted, which are the
same ones used in Ref.~\cite{Tevadetect1} (and also, for comparison,
in Ref.~\cite{last})\footnote{Apart from the cut
on the missing transverse momentum (as $W\ar \ell\nu$) ${p\Dir}_T>20$
GeV, which has not been applied here.}.
\begin{itemize}
\item For the triggered lepton $\ell$ we require 
transverse momentum $p_T^\ell>20$ GeV, pseudorapidity $|\eta^\ell|<2.5$,
and isolation from $b$-jets or partons $\Delta R_{b-jet,\ell}>0.7$.
\item Acceptance of events requires these to contain exactly two $b$-jets
(or partons) with $p_T^{b-jet}>15~\GeV$, pseudorapidity
$|\eta^{b-jet}|<2.0$, and separation $\Delta R_{b-jet,b-jet}>0.7$.
\item Moreover, events are accepted if they do not contain any
additional jet with $p_T^{j}>30$ GeV
and no more than one additional jet with 
$p_T^{j}>15$ GeV, all of them with $|\eta^{j}|<4.0$.
\end{itemize}

We first study the $WH$ case, for which we 
assume the $b$-tagging efficiency to be $\epsilon_b=50\%$ (for one
$b$)
whereas the rejection factor against non-$b$ jets is taken to be $R=50$ 
(the combination that seems to give higher significances \cite{last}). 
The backgrounds to the channel $WH\ar  \ell b\bar b X$ that
we have considered here are: $WZ$, $W\bbbar$, $q\bar q'\ar
t\bar b$, $t\bar t$, $qg\ar t\bar b q'$, $Wjb$  and $Wjj$ (see 
Ref.~\cite{Tevadetect1}).
The dominant irreducible backgrounds are $Wb\bar b$ and, especially
for $\MH$ near $\MZ$, $WZ$ with $Z\ar b\bar b$. The dominant
reducible noise comes from $Wjj$ production, via $Wgg$, $Wgq$
and $Wq\bar q$ partonic events\footnote{We checked that our codes
reproduce the numbers given in Ref.~\cite{Wjj}.}, in which
the two jets are misidentified as $b$-quarks, and  
$Wjb$ (via the partonic production
$Wqb$), for which this happens for one jet only.
Top-antitop production and decay $t\bar t\ar b\bar bW^+W^-$ with one
$W$ missed is another source of $\ell b\bar b X$ (reducible) events.
Its control requires coverage
of leptons to small $p_T$ and of jets to high rapidity (see cuts
discussed above):
this in order to reduce the probability of missing a $W$.
Moreover, in order to recognize such kind of events one can 
look for 
a $W$ via the observation of
an additional  charged lepton, of a jet with  $p_T>30$ GeV or
of two jets with $p_T>15$ GeV \cite{Tevadetect1}.
Pair production is not the only source of top quarks at the LHC.
Other important reactions that constitute a background to $WH$
are via the single top production: i.e., $Wg$ fusion (where the
`initial' $W$ is radiated off an incoming line of quarks), which gives
the final state $tbq\ar b\bar b qX$, and electroweak
fusion (into a $W$) of a $q\bar q'$-pair. Whereas the former is a
reducible background the latter is an irreducible one (as after
top decay it yields the final state $Wb\bar b$ with no additional
particle). In the first case, one can reject background events by
asking that 
the additional jet from the outgoing quark has $p_T$ less than 30 GeV.  

Figs.~3a--4a show the differential distributions of the
signal and of the sum of the various
background components in the invariant mass $M_{b\bar b}$.
We have collected the events rates in bin of 10 GeV 
around the Higgs mass (shown for 80, 100 and 120 GeV),
at the values of CM energy of $\sqrt s_{pp}=10$ and 14 TeV, respectively,
with $m_t=175$ GeV. The factor $\varepsilon$ 
indicates that the above mentioned $b$-tagging performances are included.
Tab.~II collects the event rates for all the combinations
of $\MH$ and $\sqrt s_{pp}$, in a window of 30 
GeV \cite{Tevadetect1} around
the Higgs masses, for 10 inverse femtobarns of integrated luminosity.
Figs.~3b and 4b show the various background contributions separately
(without any efficiency and/or rejection factor included).

By looking at Tab.~II one can notice how significances are
practically the same at both the energies 10 and 14 TeV, whereas
signal rates differ, in general, by $\approx35\%$.
Chances of Higgs detection are larger for a light Higgs,  
as space phase effects are important.
This situation is the opposite with respect to the channel $H\ar
\gam\gam$, where the significance of the signal grows with the
increase of the Higgs mass (see Tab.~I).
Therefore, the $H\ar b\bar b$ and $H\ar \gam\gam$ channels seem to obey
a sort of complementarity, for which the former might be the best
way to probe the Higgs mass region $80~\GeV\Ord\MH\Ord100~\GeV$,
whereas the latter is better (at high luminosity) for 
$100~\GeV\Ord\MH\Ord130~\GeV$. Therefore, our results qualitatively agree
with those given in Ref.~\cite{last}. As clearly explained there,
however, one has to remember that
the quality of the Higgs signal extraction depends less
crucially on the luminosity and on the detector performances  
in the second case. We refer the reader to Ref.~\cite{last} for a
fuller discussion about this point.

In addition to the associated production with a $W$, also the 
$q\bar q,gg\ar t\bar t H$ mechanism has been proposed as a viable channel
to search for $H\ar b\bar b$ decays \cite{btagg}, with
the $t\bar t$-pair decaying inclusively to $\ell\nu_\ell X$
($\ell=e,\mu$, as usual). Its production cross
section for an intermediate mass Higgs and $m_t\approx$ 175
GeV is of the same order as the $WH$ one (even bigger for a heavier
Higgs), see Figs.~5 and 6 of Ref.~\cite{signals}.
However, the corresponding final state is very complicated, since
it consists of four $b$-quarks: $t\bar tH\ar b\bar b b\bar b
\ell\nu_\ell X$. The signal would appear as a peak in the invariant mass 
$M_{b\bar b}$, above the combinatorial background due to the signal
itself and above the proper background processes.
Clearly, the possibilities of disentangling
the signal in this channel 
depend on the $b$-tagging performances of the LHC detectors
even more than in the $WH$ case. In particular, it has
been shown that  a high efficiency of $b$-tagging is more important than
a large rejection  of non-$b$ jets \cite{last}. 

We concentrate here on the search strategy that selects three
$b$-quarks out of the four in the final state,
as in general (except
for values of $\epsilon_b$ much greater than 50\%) the signal rates for
four tagged $b$-jets are too low \cite{last}. 
The backgrounds to the signature $b\bar b b W X$ (where the
second $b$ can be either a quark or an antiquark) in the region
80 GeV $\Ord \MH\Ord$ 140 GeV are given by the
final states $t\bar t Z$ (which is
irreducible), $t\bar t$, $t\bar t
{j}$, $W{jjj}$ and $t\bar bq$ production\footnote{The
background  $q\bar q,gg\ar t\bar t Z$ has been simulated by using the  {\tt
FORTRAN} code 
adopted in Ref.~\cite{primo}.} (which are reducible) \cite{ATLAS,last}. 

The acceptance cuts we applied are the same of the $WH$ case (for three
$b$'s/jets now and apart
from the requirement $p_T^{j}>30$ GeV on the $q$-jet of 
the $t\bar b q$ process,
which has now been dropped). Also the selection of Higgs masses we
considered here is the same, i.e., $M_H=80, 100$
and 120 GeV. Numbers are given in Tab.~III
for two combinations of $\epsilon_b$
and $R$ and for $\sqrt s_{pp}=10$ and 14 TeV.
The combined probability of picking three $b$'s out
of four in the case of the signal is here included.

In the case of $\ttbar H$ production, with respect to the case of $WH$
associated production, the chances of Higgs detection via the
$b\bar b$ decay channel are largely reduced 
(compare the significances in Tabs.~II and III). The
situation is
slightly more optimistic at 14 TeV than at 10 TeV, as at
lower energy the number of
Higgs events is reduced by approximately a factor of two, making 
the low signal rates a serious problem. Significances are also 
generally larger for lighter Higgs
masses. Notice however that these are 
the same characteristics of the $H\ar b\bar b$ signal via $WH$ production. 
Therefore, even though the $t\bar t H$ channel could not probably
constitute a serious candidate to Higgs discovery on its own,
nevertheless, if considered together with the $WH$ channel, it should
enhance the number of Higgs signals.
In addition, the irreducible background in $\ttbar Z$ events is here
quite small with respect to the signal $t\bar tH$, in the region 
$M_H\approx M_Z$, whereas this is not the case for $WZ$ and $WH$
\cite{last}. Therefore, if the Higgs mass was 
`around' the $Z$-pole, the combination $WH+t\bar tH$ in the channel
$H\ar b\bar b$ should allow for Higgs detection, 
provided that a high $b$-tagging efficiency can be
achieved, in order to remove the reducible background due to $Z\ar jj$
decays \cite{ATLAS,last}.

\subsection{Search for $H\ar ZZ^*\ar 4\ell$}
\label{sec:4l}

The four-lepton channel  $H\ar Z^{(*)}Z^{(*)}
\ar 4\ell$ with  $\ell=e$ or $\mu$ (i.e.,
$H\ar e^+e^-e^+e^- + \mu^+\mu^-\mu^+\mu^-$)\footnote{For simplicity
we neglect here the mixed case $H\ar e^+e^-\mu^+\mu^-$. However,
its inclusion in the simulation would not change the main conclusions
obtained in the case of identical leptons \cite{DellaNegra}.} 
provides the best chances for
Higgs detection over a substantially large portion of the Higgs mass
range, between $\approx 130$ and $800$ GeV. Because of this and
because its signature is relatively clean (especially 
if compared to the difficulties
encountered for the $\Hphph$ and $\Hbb$ cases),
it is often nicknamed the `gold-plated channel'.

If 130 GeV $\Ord \MH\Ord 2M_Z$, when one of the two $Z$-boson is
off-shell (i.e., $H\ar ZZ^*$), event rates are generally small 
(because of the still suppressed BR into the vector-pair) and 
the backgrounds can be dangerous (since the constraint $M_{\ell^+\ell^-}
\approx M_Z$ can be applied to only one of the lepton pairs).
However, in this mass range the Higgs width is still quite small (i.e.,
$\Gamma_H\ll 1$ GeV), therefore, provided that high enough lepton 
energy/angle resolution can be achieved, a substantial reduction
factor on the background should be possible (in fact, the rates for the latter
are directly  proportional to the resolution itself).
Also geometric and kinematical acceptances for leptons are crucial in
this case \cite{CMS14}.

In the mass region $\MH\Ord 2M_Z$ the main backgrounds to the
four-lepton signal of the \sm\ Higgs boson come from $\ttbar\ar \bbbar W^+W^-$, 
$Z\bbbar$ and $ZZ^*$ production.
In the first two backgrounds two of the leptons come from
semi-leptonic $b$-decays, whereas the remaining two from the decays
of the massive vector bosons (i.e., $W^+W^-\ar \ell^+\ell^- X$ and
$Z\ar \ell^+\ell^-$). We approximated the semileptonic 
branching fraction of the $b$-quark, i.e., $BR(b\ar \ell\nu X)$
where $\ell$ indicates a generic lepton, via the
the  one into $c$-quarks only
(neglecting then the contribution coming from the decay
$b\ar u\ell\bar\nu_\ell$),  
since the ratio between the Cabibbo-Kobayashi-Maskawa
coefficients entering in the two decays is  
$|V_{ub}/V_{cb}|=0.10\pm0.03$ \cite{DATA}. In computing
the semileptonic BR of the $b$ into $c$-quarks we have used the formula
\cite{DATA}
\be\label{bleptonic}
BR(b\ar c\ell\bar\nu_\ell)=\Gamma(b\ar c\ell\bar\nu_\ell)\tau_b=
\frac{G_F^2m_b^5}{192\pi^3}\beta(m_c,m_b)|V_{cb}|^2\tau_b, 
\ene
where $\Gamma(b\ar c\ell\bar\nu_\ell)$ is the partial  
width, $\tau_b$ the $b$-lifetime ($\approx 1.3$ ps, from the average
over $B$ hadrons)
and $\beta(m_c,m_b)=\sqrt{1-\frac{4m_c^2}{m_b^2}}$ the phase
space factor, with
$|V_{cb}|=0.043\pm0.007$
\cite{DATA} (the values for $G_F$, $m_c$ and $m_b$ were given previously).
Therefore,
the numerical value of $BR(b\ar c\ell\bar\nu_\ell)$ is approximately
$9\%$.  
As $ZZ^*$ is an intermediate stage of the 
production and decay chain $H\ar ZZ^*\ar 4\ell$, the third 
background is irreducible.

Signal rates have been generated by using the mechanisms of gluon-gluon
and vector-vector fusion production only \cite{Nigel},
as $t\bar tH$ and $WH/ZH$
are generally one order of magnitude smaller in the considered mass interval.
As usual, signals 
and backgrounds have been generated using exact tree-level matrix element
computations\footnote{We have checked 
that our results reproduce those obtained in Ref.~\cite{ggZbb} for the
subprocess $gg\ar Z b\bar b $, 
initiated by gluon-gluon fusion. In the simulations, we have used
the code already employed for the paper Ref.~\cite{primo} (which
includes also $q\bar q$-initiated subprocesses).}, 
apart from the subprocess $gg\ar ZZ^*$ (through a higher
order loop of quarks) that has not been calculated yet and that
we have simulated by multiplying the $q\bar q\ar ZZ^*$ rates by a
factor 1.3 \cite{ggZZ}.

The strategy usually adopted in order to select the four-leptons
events $H\ar ZZ^*$ out of the above backgrounds is the following.
In the case $\ell=e$, one requires:
\begin{itemize}
\item the most energetic electron must have $p_T^{e_1}>20$ GeV, the
second one $p_T^{e_2}>15$ GeV, whereas for the remaining two 
the transverse momentum must be greater than 5 GeV for both.
The rapidity of all four must be less than 2.5 (in absolute value). 
\end{itemize}   
In the case $\ell=\mu$, requirements are of the same type:
\begin{itemize}
\item the most energetic muon with $p_T^{\mu_1}>20$ GeV, the
second one with $p_T^{\mu_2}>10$ GeV, whereas the remaining two $\mu$'s
must have   $p_T^{\mu_{3,4}}>5$ GeV, all in the rapidity range $ 
|\eta^{\mu} |<2.4$.
\end{itemize}   
In both cases $\ell=e$ and $\ell=\mu$ the invariant mass cut
\begin{itemize}
\item $|M_{\ell^+\ell^-}-M_Z|<10 $ GeV is finally applied. 
\end{itemize}   
\noindent
We notice that the latter constraint is extremely successful in rejecting the
huge background coming from $\ttbar$ production, whereas it does not act
on $Zb\bar b$ and $ZZ^*$ production. However, the rates of the latter two 
processes are largely reduced by the separation criteria.
For additional background suppression we in fact required (see
\cite{CMS}) isolation on any three of the four leptons, such that
\begin{itemize}
\item there is no track with $p_T^\ell>2.5$ GeV within a $\Delta R$ cone of 
radius 0.2 around the lepton direction.
\end{itemize}   

Our numerical results are given in Tab.~IV 
for the channel $4\ell X$ and
in Tab.~V for the case $4\mu X$, where the number of
signal ($S$) and background ($B$) events is presented, together
with the corresponding significances ($S/\sqrt B$), 
for the value of integrated
luminosity ${\cal L}=100$ fb$^{-1}$.
We show results for the selection of Higgs masses $\MH=130,150$ and
170 GeV.
Significances are given for a window of 2, 4 and 6 GeV around $\MH$,
the latter two in round and squared brackets, respectively.

The behaviour of the signal cross sections (in both channels
$4\ell X$ and $4\mu X$) strongly reflects the characteristics of the 
branching ratio of the Higgs into two $Z$-bosons (see Fig.~1 in
Ref.~\cite{signals}). In fact, the highest rates occur for $M_H=150$ GeV,
value that corresponds to the local maximum in the $BR(H\ar
ZZ^*)$, whereas smaller rates occur for $\MH=130$ and 170 GeV,
the latter value  corresponding to the local minimum in the 
$ZZ$-branching ratio (due to the opening of the real $WW$-threshold).   
We verified that the dependence on $m_t$ for the signal is very
weak (see also Figs.~5a--b), such that the total significances are
practically unaffected by changes in $m_t$ (we varied the top mass in
the $t\bar t$ background accordingly).

Figs.~5--8 show the distribution in the invariant mass
of the 4 leptons/muons. They give account also of the shape of the background.
This is dominated by the $t\bar t$ component in the region
$M_{4\ell,4\mu}\approx 130$ GeV, 
whereas the local maximum around $M_{4\ell,4\mu}=190$ GeV is 
due to the opening of the threshold for the production of two real
$Z$'s in the background $pp\ar ZZ$. 
The background from $pp\ar Z b\bar b$ is never dominant (see also
\cite{ATLAS,CMS}).

For the case of the $4\ell X$ channel, if the high luminosity option
can be achieved, it should be possible to observe the
Higgs boson in the intermediate mass range after one year or so of running,
for all combinations of $\MH$ and $\sqrt s_{pp}$.
Particularly favourable is the case of $\MH=150$ GeV, which should be
resolved even for 10 fb$^{-1}$, both at 10 and 14 TeV of CM energy. 
For the lower luminosity option the case $\MH=170$ GeV would present
problems, independently of the mass resolution that could be achieved,
whereas for $\MH=130$ GeV, this happens if $\Delta M\OOrd2$ GeV,
especially at 10 TeV.  

The four muon case reflects more o less the same characteristics
of the total $4\ell$ case.
Rates are obviously smaller than in the $4e+4\mu$ case, however,
acceptance  criteria generally favour muon cross sections more than
electron ones, 
both for signals and backgrounds. But 
whereas for the signal
this happens  with differences of just a few percents,
in the case of the backgrounds the muon decays
of the top quarks (after the above cuts) are
about 40\% larger than those into electrons, at fixed
$\sqrt s_{pp}$ and $m_t$. Since top production is in general
the dominant background,
the final significances for the muon case are typically smaller  
than  the ones of the $4\ell$ case. Therefore, even more in this case,
at low luminosity, Higgs with masses far away from
the local maximum in the $BR(H\ar ZZ^*)$ at $M_H=150$ GeV, such as
at 130 and 170 GeV, would be overwhelmed by the background, unless 
a really good muon resolution can be achieved (the latter is in fact
the chance that induces to consider the four muon decay separately).
For the case $M_H\approx170$ GeV, the recourse to the channel
$H\ar W^{(*)}W^{(*)}\ar \ell^+\nu_\ell \ell'^-\bar\nu_{\ell'}$,
advocated in Ref.~\cite{dittdrei}, might then be providential.

We would like here to conclude this subsection by
stressing the importance of the isolation cut on three of the
leptons, as this can greatly reduce the $t\bar t$ background (of a factor
$\approx5$) and the $Zb\bar b $ one (of a factor $\approx2$). At the same
time almost 95\% of the $4e$ and $4\mu$ signals pass this cut. 

\section{The heavy mass range}

\subsection{Search for $H\ar ZZ\ar 4\ell$, with  $\ell=e$ or $\mu$}
\label{sec:4lheavy}

In the case of heavy Higgs mass, when  the on-shell decay $H\ar
ZZ$ can take place, the only significant background (after applying
the invariant mass cut $M_{\ell^+ \ell^-}\approx M_Z$ on both the 
lepton-antilepton pairs) is the continuum $ZZ$ \cite{ATLAS,CMS}.
Therefore, we concentrate on this background only. Here, $p_T$ and
$\eta$ selection cuts on the four $\ell$'s are the same as in the
$ZZ^*$ channel for an intermediate mass Higgs. 
The sole difference is that in the heavy mass range we drop the
requirement of isolation on three leptons, as it was 
previously implemented in view of
reducing the $t\bar t$ background, which is here negligible compared
to the $ZZ$ contribution.

The invariant mass $M_{4\ell}$ for signal and
background is shown in Figs.~\Hfourlaheavy--\Hfourlbheavy, 
for a CM energy of 10 and 14 TeV, respectively.
The values for $\MH$ here considered are 300, 500 and 700 GeV. In particular,
Figs.~9a and 10a clearly show the feasibility of Higgs
detection for a mass around 300 GeV, both at 10 and 14 TeV.
For example,
the number of signal and background events, for ${\cal L}=100$
fb$^{-1}$, at $\sqrt s_{pp}=10$(14) TeV, is 214(408) and 123(102), 
respectively, in a 70 GeV window around the Higgs peak 
(i.e., $|M_{H}-M_{4\ell}|<35$ GeV, since the total Higgs width 
for $\MH=300$ GeV is equal to $\approx8.4$ GeV). 
Even the case of low luminosity (i.e., ${\cal L}=10$ fb$^{-1}$)
should allow one to reveal the Higgs scalar, after only one year of running. 
Higgs signals would appear as a broad Breit-Wigner resonance on top of
a decreasing background (with increasing $\MH$), 
which has a maximum at $M_{4\ell}=2M_Z$.
The case of a Higgs scalar with mass around 500 GeV also allows for
Higgs detection, as can be seen by looking at Figs.~9b and 10b.

Things are more complicated when one considers larger Higgs masses, like
for example $\MH=700$ GeV
(Figs.~9c--10c). For such a value one looses the concept of 
resonance, as the Higgs particle has a width comparable to its mass, i.e.,
$\Gamma_H\approx 187$ GeV, and the characteristic Breit-Wigner peak
disappears. In this case the shapes of signal and backgrounds are
no longer easily distinguishable, and one has to be extremely careful
in normalising the distributions. For the first stage energy $\sqrt
s_{pp}=10$ GeV, signal and background are hard to recognize (see
Fig.~9c). 
At $\sqrt s_{pp}=14$ TeV prospects are more optimistic, as here the 
maximum of the Higgs distribution at 700 GeV is 
a factor of three over
the $ZZ$ background (see Fig.~10c). In general, at higher CM energy,
Higgs detection should then be 
feasible both at low and high collider luminosity.

\subsection{Search for \Hllnn\ ($\ell=e,\mu$ and $\ell'=e,\mu,\tau$)}
\label{sec:llnn}

The advantage of the $H\ar \ell^+\ell^-\nu_{\ell'}\bar\nu_{\ell'}$ 
decay channel
with respect to $H\ar 4\ell$ is that it has a BR which is six times larger.
The disadvantage is that, because of the presence of two neutrinos
in the final state, invariant mass peaks around the actual Higgs mass value
cannot be reconstructed. Therefore, this channel prevents from accurate 
studies of the Higgs parameters: e.g., the width $\Gamma_H$.
The distinctive signature is two high $p_T$ leptons (from one
$Z$-decay) and high missing (transverse) energy $E_{T}^{{miss}}$ (from the 
other $Z$).
The main backgrounds to the two-lepton-two-neutrino channel are the
continuum production of two massive vector bosons ($WZ$ and $ZZ$) plus
the top-antitop production ($t\bar t$) and $Z$ + jets. After the event
selection cuts, the dominant background to the signal 
is the irreducible one $pp\ar ZZ$, 
whereas the other three give smaller
contributions  \cite{CMS2223,ATLAS3839}.

For $\MH\OOrd500$ GeV
Higgs signals appear as broad Jacobian peaks in the two-lepton
transverse momentum distribution. To select them
we adopt the following procedure \cite{CMS}. We require:
\begin{itemize}
\item $E_T^{{miss}}>100$ GeV (which strongly suppresses the $Z+$
jets
background)\footnote{In fact, for $E_T^{{miss}}<150$ GeV such
process dominates over the others, as the jets either can escape the
detector acceptance region or can be mis-measured in the calorimeter (see
\cite{ATLAS}).};
\item $p_T^\ell>20$ GeV, $|\eta^\ell|<1.8$ and $p_T^{\ell\ell}>60$
GeV;
\item $|M_{\ell^+\ell^-}-M_Z|\le6$ GeV (which strongly reduces the reducible
backgrounds);
\item no jet with $E_T^j>150$ GeV within $|\eta^{{j}}|
\ge1.8$; 
\item no back-to-back jets with the leptons (i.e., the cosine of the
angle between the momentum of the jets and of $\ell^+\ell^-$ should be
greater than $-0.8$).
\end{itemize} 

Figs.~11 and 12 show the distribution on the transverse momentum of the
lepton pair $\ell^+\ell^-$, after all cuts, for $M_H=500$ GeV, 
at $\sqrt s_{pp}=10$ and 14 TeV, respectively,
for signal and background.
The dependence on the collider CM is such that at 14 TeV one
gets significances which are a factor 1.5 greater than
at 10 TeV, with the number of events doubled.
Assuming an integrated luminosity of $10$ fb$^{-1}$, one obtains
(in the window, say, 
$|p_T^{\ell\ell}-225~{\mbox{GeV}}|\le25$ GeV) 
31.7(12.6) signal(background) events at 10 TeV. 
The corresponding numbers at $\sqrt s_{pp}=14$ TeV are 67.6(19.4). 
Note that, in order to suppress the $t\bar t$ background, the additional cut
$|M_{Z}-M_{\ell^+\ell^-}|\le6$ GeV has been applied.

From Figs.~11--12 is then clear that both
at 10 and at 14 TeV extremely good chances of Higgs detection
(for masses, say, $M_H\approx500$ GeV) exist\footnote{We have 
also checked that our conclusions
about heavy Higgs detection are essentially
unmodified if one varies the value of $\MH$ in the region
400--700 GeV, although for higher values of the Higgs mass things
are complicated by the fact that the background has a similar shape,
whereas
for lower values the peak at soft $p_T^{\ell\ell}$ of the background
largely covers the Jacobian peaks of the Higgs boson. 
A different optimisation of the cuts could however extend the
mass range
of observability of the Higgs boson via 
the channel $H\ar \ell^+\ell^-\nu_{\ell'}\bar\nu_{\ell'}$.}, 
by resorting to the two-lepton-two-neutrino channel, already for 10
inverse femtobarns. Therefore,
different values of $\sqrt s_{pp}$ do not modify the
search strategies of the Higgs boson via this channel and have a 
negligible impacts on the significances in the signal vs. background
analysis.

The $H\ar \ell^+\ell^-\nu_{\ell'}\bar\nu_{\ell'}$ 
channel has been claimed to give
good chances for Higgs detection also for higher values of $M_H$,
in the range 700 GeV--1 TeV. However, as in this mass region 
the Higgs production cross sections via the vector-vector fusion
mechanism (which are comparable to the ones
obtained via gluon-gluon fusion, see Figs.~5a--b in Ref.~\cite{signals}) 
are strongly affected by unitarisation corrections to the 
longitudinal $V_LV_L$ 
scattering (where $V=W,Z$) \cite{breaks}, we feel our approach 
(via tree--level perturbative amplitudes) inadequate to treat
conveniently such delicate region of the Standard Model. This is why 
we preferred to restrict our attention to lower values of 
$\MH$.    

\section{Summary and conclusions}

This paper has been devoted to study the phenomenology of the Higgs
boson of the Standard Model at the Large Hadron Collider, the next
generation $pp$ collider at CERN, by comparing signal-to-background rates
and by computing the corresponding significance factors.
We have used in our analysis one of the most recent parton distributions 
as well as the
most updated values of the input parameters of the \sm.
We have continued here the work begun in Ref.~\cite{signals}, where up-to-date
Higgs cross sections, branching ratios and event rates were presented.

The signatures by which Higgs detection is most promising at the LHC
are the following:
\begin{itemize}
\item $t\bar tH+WH\ar \ell\nu_\ell\gamma\gamma X$, 
in the intermediate mass range;
\item $WH\ar \ell\nu_\ell b\bar
bX$ and $t\bar t H
\ar b\bar b b\bar b  \ell\nu_\ell X$, 
in the intermediate mass range;
\item $H\ar
Z^{(*)}Z^{(*)}\ar \ell^+\ell^-\ell'^{+}\ell'^{-}$, where $\ell=e$ or $\mu$, 
both in the intermediate and heavy mass range;
\item $H\ar
ZZ\ar \ell^+\ell^-\nu_{\ell'}\bar\nu_{\ell'}$, where $\ell=e$ or $\mu$ and
$\ell'=e,\mu$ or $\tau$,  
in the heavy mass range.
\end{itemize}

In order to make a signal-to-background analysis as meaningful as
possible, we have {\sl consistently} evaluated both 
signal and background rates at leading
order. This has been done because the most part of the backgrounds 
have been so far calculated (and here independently re-computed)\footnote{It 
made
exception in this context only the usage of next-to-leading order parton
distributions and of $\alpha_s$ at two-loops.}  at
lowest order only. 
Results have been given for the two planned CM 
energies of 10 and 14 TeV, 
and discussed on the basis of the expected integrated
luminosity,  10 and/or 100 inverse femtobarns $per$ $annum$.

What has been assessed here is the substantial 
complementarity of the decay channels $\gam\gam$ and $b\bar b$ 
in the intermediate mass range, as already noticed in
Ref.~\cite{last}.
In fact, the latter is the most promising in covering the mass
region 80 GeV $\Ord\MH\Ord$ 100 GeV, whereas the former is better if
100 GeV $\Ord\MH\Ord$ 130 GeV. Furthermore,
the option of high luminosity would make the detection much more feasible
for the Higgs in the di-photon channel whereas it
would reduce the chances via
the $b\bar b$-channel, as in this case
both the lepton trigger threshold and the
jet $p_T$ threshold would certainly have to be raised in order to
compensate the increased trigger rates (see discussion in
Ref.~\cite{last}),  thus reducing even more
the already low rates of both the $WH$ and $\ttbar H$ signals. Finally,
a sort of complementarity
exists also among the latter two production mechanisms.
In fact, on the one hand, the $t\bar t H$ signal has a
quite small irreducible background in $t\bar t Z$ events, whereas
this is not the case for $WZ$ events, which have rates competitive with
the ones of the $WH$ signal, if 
$M_H\approx M_Z$. On the other hand, since the $t\bar tH\ar
 \ell\nu b\bar b (b\bar b) X$ signal is affected by combinatorial background,
whereas $WH\ar \ell\nu b\bar b X$ is not, the identification of Higgs
signals for $M_H\not\approx\MZ$ is easier in the second case. 
All of this is valid for both  the
energy stages of the LHC, 10 and 14 TeV.
In the case of poor detector performances, things would be instead quite 
complicated.
Finally, as in general 
(for these two channels) the difference between the signal significances
at 14 and 10 TeV is a factor of approximately 1.3--1.5,
it  should be considered that one should run the
machine at 10 TeV twice the time that is needed at 14 TeV to get to
the same threshold of observability of an intermediate mass Higgs.
These conclusions, however, largely rely on the fact that the high
luminosity option could be achieved in reasonable time for
the $\gam\gam$-channel and that the expected
performances of the LHC detectors (especially in photon angular and energy
resolution for the $\gamma\gam$ case and in $b$-tagging for the $b\bar b$ one)
could be confirmed in practice.

For $130~\GeV\Ord\MH\Ord 2M_Z$, the below threshold decay channel $H\ar ZZ^*$
gives good chances of Higgs detection. For reasonable mass
resolutions of the detectors (see Refs.~\cite{ATLAS} and \cite{CMS}),
and  if the high luminosity option
can be achieved, it should be possible an early observation of the
Higgs boson, 
for both
values of the collider energy, 10 and 14 TeV. As the peak in
the $ZZ^*$ branching ratio (i.e., below the threshold for two real
$Z$'s) occurs for $\MH=150$ GeV, the best chances of disentangling the
signals would occur for Higgs masses around this value, particularly
for 10 fb$^{-1}$ of luminosity. 
For such a value of ${\cal L}$ the case $\MH=170$ GeV would present
serious problems, independently of the mass resolution,
whereas for $\MH=130$ GeV this happens if $\Delta M\OOrd2$ GeV,
especially at 10 TeV.  
The case of four muons reflects more o less the same characteristics
of the total $4\ell$ case.
 
For a heavy Higgs, with $\MH\OOrd 2M_Z$, detection should be
guaranteed up to values of $\approx700$ GeV, via the four-lepton
channel. For Higgs heavier than, say, 500 GeV, the channel
$H\ar \ell^+\ell^-\nu_{\ell'}\bar\nu_{\ell'}$ 
will give additional/alternative possibilities.
The only difficult conditions would be in the case of a very heavy 
Higgs boson (around 700 GeV), since in this case the corresponding decay
width is very large (of the same order as the mass)
and the particle cannot be considered a resonance
any longer, such that the shape of the distribution in the invariant
mass of the four leptons is the same as the one from the background
processes. This would especially be the case at $\sqrt s_{pp}=10$ TeV.
In such conditions, the recourse to the
channel $H\ar \ell^+\ell^-\nu_{\ell'}\bar\nu_{\ell'}$ could turn 
out to be very useful, as, on the one hand,
the corresponding branching ratio is six times larger
than that into four leptons (so event rates would be much larger, thus
allowing for larger significances) whereas, on the other hand, the signal
does not suffer so strongly from broadening effects, since this 
latter is not
a Breit-Wigner peak but a Jacobian one due to the kinematics of the Higgs. 

According to the results of our analysis,
we conclude that, at the latest after a few
years of running at 14 TeV (the long running compensating the possible reduced
performances of the detectors), the LHC might be able to {\sl definitely}
asses the correctness of the Minimal Standard Model, or to rule out 
this with certainty. Moreover, if the expected efficiency of the
detectors will be achieved in time, then it might be
possible to say something decisive about the Standard Model already
after the first energy stage, around 2005. 
 
Before closing, we would like to remind the reader
two decisive aspects of the analysis
that we have been carrying out, which indicate how the results we have
obtained should be taken as an {\sl indication} of the Higgs 
discovery potential 
of the LHC at its two planned stages of energy and in dependence of
the integrated luminosity, 
more than as a complete analysis of
signals, backgrounds and search strategies. 

First, the analysis has been done {\sl exclusively} but {\sl consistently} 
at {\sl parton} level. Apart from the (standard) integration procedure of the
differential cross sections, no recourse to any Monte Carlo technique
and/or event generator has been done.
Only exact {\sl perturbative} 
matrix element computations  have been
performed. Hadronization and fragmentation effects, jet clustering procedures, 
initial and final QCD and QED radiation have not been
contemplated, and no
special effort has been done in simulating the realistic detector performances
(smearing and pile-up effects, finite efficiency in lepton
identification, conversion
losses for photons, etc ...), other than adopting the usual selection
criteria, which can be largely implemented also at parton level
(transverse momenta and (pseudo)rapidity of leptons and jets, 
angular separations, missing energy, etc ...).
Therefore, a systematic uncertainty due to the unavoidable
differences between parton-level and jet-level procedures 
come with our results. The interplay between these two approaches 
has been carefully investigated and quantified
in Ref.~\cite{last}, in the case of the $b\bar b$ channel. 
In general, the differences are larger for processes with a complex
and various final state topology (such as the one involving top production
and decay), and they mainly concern the treatment of hadronic final states, 
more than the one of leptons and neutrinos.  
This observation can be safely extended to involve also the case of the
other Higgs decay channels (and relative backgrounds).
However, we do not expect the inclusion of all these non-perturbative
aspects to wash out
the main results that have been assessed here, also because many of the
Higgs signatures studied in this paper were indeed 
non-hadronic (i.e., $H\ar \gam\gam, 4\ell, 2l2\nu$). 
In contrast, many of the inconsistencies that often occur in such kind
of analyses (for example,
when NLO order computations are compared to LO results,
when out-of-date parton distributions are used, when parton shower
approximations are substituted to or interfaced with exact
computations, etc ...) are here totally removed.

Second, it has to be remembered that higher order rates for the most
part of the backgrounds have not been computed yet. Therefore,
especially when proceeding to a signal-to-background
analysis at lowest order
(thus avoiding the use of $K$-factors for the signals, as it
has been done here), an overall uncertainty due to the lack of
knowledge of next-to-leading order corrections remains. 
This has been estimated, for example, 
 for the case of the $WH$ signal and corresponding
backgrounds,  in Ref.~\cite{KfacVH}, where it was shown that their 
inclusion does not change the final results significantly. 
Once all the rates needed for a self-consistent signal-to-background
analysis will be available at next-to-leading order, very accurate 
predictions about the prospects of Higgs detection and study  at
the LHC will be given\footnote{In this respect, we inform the reader
that the calculation of the 
complete QCD corrections at NLO to the $pp\ar t\bar tH$ 
production channel are now well under way \cite{ioeben}.}. 
It is hard to think, in fact, that
next-to-next-to-leading corrections can significantly
 modify
the conclusions obtained at the preceeding order. Possibly, an important
exception could be the case of the gluon-gluon fusion process, $gg\ar H$,
for which these could well be large, as the next-to-leading rates
differ by $\approx100\%$ from the leading order ones \cite{signals}.
Because of the crucial role that this production channel has
at the LHC, it is important that such corrections
are soon investigated. 

Finally, we remind the reader that the top mass is no longer 
a significant source of theoretical error in Higgs searches, as this particle
has been finally identified at FNAL and its mass measured rather accurately, 
i.e., $m_t=175\pm6$ GeV. Certainly, there is a residual uncertainty in the
predictions involving the production of (virtual or real) $t$-quarks due
to the existing experimental error. However it has been shown (see Fig.~6
in Ref.~\cite{signals}) 
that its impact on the Higgs production rates is generally small, yielding
differences in the cross sections which are generally below $8-9\%$. 
We have verified that similar effects also occur in the case of the 
top backgrounds. The only
exceptions are the NLO rates for the signal $gg\ar H$ (via
top loops), for which such differences can be as large as 30\% (near
the threshold $M_H\approx2m_t$). However, this has  a limited
phenomenological relevance, as for that value of $M_H$ the Higgs signals
are well above the backgrounds (see, e.g., Fig.~9a and 10a and compare to
the rates given in the two tables of Ref.~\cite{signals}). Furthermore,
one should remember that by the time
the LHC comes into operation the top quark mass will be known with much better
precision so that the accuracy in the prediction of top cross sections
will be even higher than that discussed in this paper.

\section{Acknowledgments}

This work is supported in part by the
Ministero dell' Universit\`a e della Ricerca Scientifica, 
by the UK Particle Physics and Astronomy Research Council and by the EC 
Programme ``Human Capital and Mobility'', contract CHRX--CT--93--0357
(DG 12 COMA). 
We are grateful to E. Maina for a careful reading
of the manuscript and to W.J. Stirling for useful suggestions.
We would also like to thank all the members of the Centre for Particle Theory 
of Durham University (and among them, in particular,
Mike Whalley, for his incomparable help in `digging' me out of 
tricky computing problems), 
where the most part of this work has been carried out, 
for hosting me during eighteen friendly and scientifically stimulating 
months.

\vfill
\newpage
\section*{Table captions}

\begin{itemize}

\item[{[I]}] Number of signal ($S$, $WH+t\bar tH$)
and background ($B$, $W\gamma\gamma+W\gamma{j}+t\bar t\gamma\gamma+
t\bar t\gamma+Z\gamma$) events, together with the significance
$S/\sqrt B$, for the signature $\ell \gam\gam X$, 
in the windows $|M_{H}-M_{\gam\gam}|<1(2)[3]$ GeV,
for ${\cal L}=100$ fb$^{-1}$ and the selection of Higgs masses
$\MH=80,100$ and 120 GeV, after the cuts mentioned in the
text, at $\sqrt s_{pp}=10$ TeV (upper section) and 14 TeV (lower section),
for $m_t=175$ GeV.
In order to suppress the background $Z\gam$ the additional cut
$|M_{Z}-M_{\ell\gam}|>5$ GeV is applied. 

\item[{[II]}] Number of signal ($S$, $WH$)
and background ($B$, $WZ+Wb\bar b+t\bar b+
t\bar t+t\bar bq+W{jj}+Wb{j}$) events, together with the significance
$S/\sqrt B$, for the signature $\ell b\bar b X$, 
in the window $|M_{H}-M_{b\bar b}|<15$ GeV,
for ${\cal L}=10$ fb$^{-1}$ and the selection of Higgs masses
$\MH=80,100$ and 120 GeV, after the cuts mentioned in the
text, at $\sqrt s_{pp}=10$ TeV (upper section) and 14 TeV (lower section),
for $m_t=175$ GeV.
The $b$-tagging performances are
$\epsilon_b=50\%$ and $R=50$.

\item[{[III]}] Number of signal ($S$, $t\bar tH$)
and background ($B$, $t\bar tZ+t\bar t{j}+W{jjj}+
\ttbar +t\bar bq$) events, together with the significance
$S/\sqrt B$, for the signature $\ell b\bar b b X$, 
in the window $|M_{H}-M_{b\bar b}|<15$ GeV,
for ${\cal L}=10$ fb$^{-1}$ and the selection of Higgs masses
$\MH=80,100$ and 120 GeV, after the cuts mentioned in the
text, at $\sqrt s_{pp}=10$ TeV (upper section) and 14 TeV (lower section),
for $m_t=175$ GeV.
The $b$-tagging performances are 
$\epsilon_b=50\%$ and $R=50$ (lower row), $\epsilon_b=70\%$ and $R=10$
(upper row).

\item[{[IV]}] Number of signal ($S$, $H+q\bar qH$) and background ($B$, 
$t\bar t+Zb\bar b+ ZZ^*$) events, together with the significance
$S/\sqrt B$, for the signature $4\ell X$, 
in the windows $|M_{H}-M_{4\ell}|<1(2)[3]$ GeV,
for ${\cal L}=100$ fb$^{-1}$ and the selection of Higgs masses
$\MH=130,150$ and 170 GeV, after the cuts mentioned in the
text, at $\sqrt s_{pp}=10$ TeV (upper section) and 14 TeV (lower section),
for $m_t=175$ GeV.
In order to suppress the background $t\bar t$ the additional cut
$|M_{Z}-M_{\ell^+\ell^-}|<10$ GeV is applied.

\item[{[V]}] Number of signal ($S$, $H+q\bar qH$) and background ($B$, 
$t\bar t+Zb\bar b+ ZZ^*$) events, together with the significance
$S/\sqrt B$, for the signature $4\mu X$, 
in the windows $|M_{H}-M_{4\mu}|<1(2)[3]$ GeV,
for ${\cal L}=100$ fb$^{-1}$ and the selection of Higgs masses
$\MH=130,150$ and 170 GeV, after the cuts mentioned in the
text, at $\sqrt s_{pp}=10$ TeV (upper section) and 14 TeV (lower section),
for $m_t=175$ GeV.
In order to suppress the background $t\bar t$ the additional cut
$|M_{Z}-M_{\mu^+\mu^-}|<10$ GeV is applied.



\end{itemize}

\vfill
\newpage
\section*{Figure captions}

\begin{itemize}

\item[{[1]}] Distribution in invariant mass of the photon-photon
pair for signal ($WH+t\bar tH$)
and background ($W\gamma\gamma+W\gamma{j}+t\bar t\gamma\gamma+
t\bar t\gamma+Z\gamma$),
giving the signature $\ell \gam\gam X$,
after the cuts mentioned in the
text, at $\sqrt s_{pp}=10$ TeV and for $m_t=175$ GeV (a).
In order to suppress the background $Z\gam$ the additional cut
$|M_{Z}-M_{\ell\gam}|>5$ GeV is applied.
The symbol $\varepsilon$ indicates that also the reduction factors
due to misidentification of a lepton or a jet for a photon
are included, in the case of the backgrounds
$Z\gam$, $t\bar t\gam$ and $W\gam{j}$. In (b) the various
background contributions are shown separately (with $m_t=175$ GeV for
top contributions).

\item[{[2]}] Same as Fig.~1, for $\sqrt s_{pp}=14$ TeV.

\item[{[3]}] Distribution in invariant mass of the $b\bar b$--pair
for signal ($WH$)
and background ($WZ+Wb\bar b+t\bar b+
t\bar t+t\bar bq+Wb{j}+W{jj}$),
giving the signature $\ell b\bar b X$,
after the cuts mentioned in the
text, at $\sqrt s_{pp}=10$ TeV and for $m_t=175$ GeV (a).
In the case of the backgrounds $t\bar t$ and $t\bar bq$ we have also
implemented the cuts indicated in the text for the additional jets 
in the final state.
The symbol $\varepsilon$ indicates that efficiencies and reduction
factors of $b$-tagging are included, both for signals and backgrounds.
In (b) the various
background contributions are shown separately (with $m_t=175$ GeV for
top contributions).
  
\item[{[4]}] Same as Fig.~3, for $\sqrt s_{pp}=14$ TeV.

\item[{[\Hfourla]}] Distribution in invariant mass of the two
lepton pairs
for the sum of signal ($H+q\bar qH$)
and background ($t\bar t+Zb\bar b+
ZZ^*$, the latter in shadowing),
giving the signature $4\ell X$,
after the cuts mentioned in the
text, at $\sqrt s_{pp}=10$ TeV and for $m_t=175$ GeV.
In order to suppress the background $t\bar t$ the additional cut
$|M_{Z}-M_{\ell^+\ell^-}|<10$ GeV is applied.

\item[{[\Hfourlb]}] Same as Fig.~5, for $\sqrt s_{pp}=14$ TeV.

\item[{[\Hfourma]}] Distribution in invariant mass of the four
muons
for the sum of signal ($H+q\bar qH$)
and background ($t\bar t+Zb\bar b+
ZZ^*$, the latter in shadowing),
giving the signature $4\mu X$,
after the cuts mentioned in the
text, at $\sqrt s_{pp}=10$ TeV and for $m_t=175$ GeV.
In order to suppress the background $t\bar t$ the additional cut
$|M_{Z}-M_{\mu^+\mu^-}|<10$ GeV is applied.

\item[{[\Hfourmb]}] Same as Fig.~7, for $\sqrt s_{pp}=14$ TeV.

\item[{[\Hfourlaheavy]}] Distribution in invariant mass of the two
lepton pairs
for signal ($H+q\bar qH$, in shadowing)
and background ($ZZ$),
giving the signature $4\ell X$,
after the cuts mentioned in the
text, at $\sqrt s_{pp}=10$ TeV, for $m_t=175$ GeV
and for $M_H=300$ (a), 500 (b) and 700 (c) GeV.

\item[{[\Hfourlbheavy]}] Same as Fig.~9, for $\sqrt s_{pp}=14$ TeV.

\item[{[\Hfourlaobese]}] Distribution in transverse momentum of the
lepton pair
for signal ($H+q\bar qH$, in shadowing)
and background ($ZZ+ZW+t\bar t+~Z+\mbox{jets}$),
giving the signature $\ell^+\ell^-\nu_{\ell'}\bar\nu_{\ell'} X$,
after the cuts mentioned in the
text, at $\sqrt s_{pp}=10$ TeV, for $m_t=175$ GeV
and for $M_H=500$ GeV.
In order to suppress the background $t\bar t$ the additional cut
$|M_{Z}-M_{\ell^+\ell^-}|\le6$ GeV is applied.

\item[{[\Hfourlbobese]}] Same as Fig.~11, for $\sqrt s_{pp}=14$ TeV.

\end{itemize}

\vfill
\newpage

\begin{table}
\begin{center}
\begin{tabular}{|c|c|c|c|}
\hline
\multicolumn{4}{|c|}
{\rule[0cm]{0cm}{0cm}
$\ell\gamma\gamma X$}
 \\ \hline  \hline
\rule[0cm]{0cm}{0cm}
$M_H$~(GeV) &  $S$  &  $B$/2 GeV  &  $S/\sqrt B$ \\ \hline\hline
\rule[0cm]{0cm}{0cm}
$~80$ &  $28.6$  &  $6.9$  &  $10.9(7.7)[6.3]$ \\ \hline
\rule[0cm]{0cm}{0cm}
$100$ &  $37.7$  &  $5.5$  &  $16.1(11.4)[9.3]$ \\ \hline
\rule[0cm]{0cm}{0cm}
$120$ &  $37.9$  &  $4.0$  &  $18.9(13.4)[10.9]$ \\ \hline
\multicolumn{4}{|c|}
{\rule[0cm]{0cm}{0cm}
$\sqrt s_{pp}=10$ TeV}
 \\ \hline  \hline
\rule[0cm]{0cm}{0cm}
$~80$ &  $44.6$  &  $9.7$  &  $14.3(10.1)[8.3]$ \\ \hline
\rule[0cm]{0cm}{0cm}
$100$ &  $59.0$  &  $8.0$  &  $20.9(14.8)[12.0]$ \\ \hline
\rule[0cm]{0cm}{0cm}
$120$ &  $60.0$  &  $6.1$  &  $24.3(17.2)[14.0]$ \\ \hline
\multicolumn{4}{|c|}
{\rule[0cm]{0cm}{0cm}
$\sqrt s_{pp}=14$ TeV}
 \\ \hline  \hline
\multicolumn{4}{|c|}
{\rule[0cm]{0cm}{0cm}
$m_t=175$ GeV}
 \\ \hline 
\multicolumn{4}{c}
{\rule{0cm}{1cm}
{\Large Tab. I}}  \\
\multicolumn{4}{c}
{\rule{0cm}{0cm}}
\end{tabular}
\end{center}
\end{table}

\vfill
\newpage

\begin{table}
\begin{center}
\begin{tabular}{|c|c|c|c|}
\hline
\multicolumn{4}{|c|}
{\rule[0cm]{0cm}{0cm}
$\ell b\bar b X$ (Events/30 GeV)}
 \\ \hline  \hline
\rule[0cm]{0cm}{0cm}
$~80$ &  $224$  &  $460$  &  $10.$ \\ \hline
\rule[0cm]{0cm}{0cm}
$100$ &  $140$  &  $323$  &  $7.8$ \\ \hline
\rule[0cm]{0cm}{0cm}
$120$ &  $~74$  &  $255$  &  $4.6$ \\ \hline
\multicolumn{4}{|c|}
{\rule[0cm]{0cm}{0cm}
$\sqrt s_{pp}=10$ TeV}
 \\ \hline  \hline
\rule[0cm]{0cm}{0cm}
$~80$ &  $299$  &  $715$  &  $11.$ \\ \hline
\rule[0cm]{0cm}{0cm}
$100$ &  $187$  &  $517$  &  $8.2$ \\ \hline
\rule[0cm]{0cm}{0cm}
$120$ &  $100$  &  $411$  &  $4.9$ \\ \hline
\multicolumn{4}{|c|}
{\rule[0cm]{0cm}{0cm}
$\sqrt s_{pp}=10$ TeV}
 \\ \hline  \hline
\multicolumn{4}{|c|}
{\rule[0cm]{0cm}{0cm}
$m_t=175$ GeV}
 \\ \hline 
\multicolumn{4}{c}
{\rule{0cm}{1cm}
{\Large Tab. II}}  \\
\multicolumn{4}{c}
{\rule{0cm}{0cm}}
\end{tabular}
\end{center}
\end{table}

\vfill
\newpage

\begin{table}
\begin{center}
\begin{tabular}{|c|c|c|c|}
\hline
\multicolumn{4}{|c|}
{\rule[0cm]{0cm}{0cm}
$\ell b\bar b b X$ (Events/30 GeV)}
 \\ \hline  \hline
\rule[0cm]{0cm}{0cm}
$M_H$~(GeV) &  $S$  &  $B$  &  $S/\sqrt B$ \\ \hline\hline
\rule[0cm]{0cm}{0cm}
$~80$ &  $46$  &  $177$  &  $3.6$ \\ 
$~~~$ &  $17$  &  $27$   &  $3.3$ \\ \hline
\rule[0cm]{0cm}{0cm}
$100$ &  $27$  &  $176$  &  $2.0$ \\ 
$~~~$ &  $10$  &  $26$   &  $2.0$ \\ \hline
\rule[0cm]{0cm}{0cm}
$120$ &  $14$  &  $165$  &  $1.1$ \\
$~~~$ &  $5$  &   $23$   &  $1.0$ \\ \hline
\multicolumn{4}{|c|}
{\rule[0cm]{0cm}{0cm}
$\sqrt s_{pp}=10$ TeV}
 \\ \hline  \hline
\rule[0cm]{0cm}{0cm}
$~80$ &  $98$  &  $335$  &  $5.4$ \\
$~~~$ &  $35$  &  $54$   &  $4.8$ \\ \hline
\rule[0cm]{0cm}{0cm}
$100$ &  $57$  &  $341$  &  $3.1$ \\
$~~~$ &  $21$  &  $51$   &  $2.9$ \\ \hline
\rule[0cm]{0cm}{0cm}
$120$ &  $30$  &  $318$  &  $1.7$ \\
$~~~$ &  $11$  &  $46$   &  $1.6$ \\ \hline
\multicolumn{4}{|c|}
{\rule[0cm]{0cm}{0cm}
$\sqrt s_{pp}=14$ TeV}
 \\ \hline  \hline
\multicolumn{4}{|c|}
{\rule[0cm]{0cm}{0cm}
$m_t=175$ GeV}
 \\ \hline 
\multicolumn{4}{c}
{\rule{0cm}{1cm}
{\Large Tab. III}}  \\
\multicolumn{4}{c}
{\rule{0cm}{0cm}}
\end{tabular}
\end{center}
\end{table}

\vfill
\newpage

\begin{table}
\begin{center}
\begin{tabular}{|c|c|c|c|}
\hline
\multicolumn{4}{|c|}
{\rule[0cm]{0cm}{0cm}
$4\ell X$}
 \\ \hline  \hline
\rule[0cm]{0cm}{0cm}
$M_H$~(GeV) &  $S$  &  $B$/2 GeV  &  $S/\sqrt B$ \\ \hline\hline
\rule[0cm]{0cm}{0cm}
$130$ &  $44.7$  &  $19.0$  &  $10.3(7.4)[6.1]$ \\ \hline
\rule[0cm]{0cm}{0cm}
$150$ &  $103.6$  &  $11.82$  &  $30.1(21.0)[17.1]$ \\ \hline
\rule[0cm]{0cm}{0cm}
$170$ &  $20.6$  &  $6.8$  &  $7.9(5.7)[4.7]$ \\ \hline
\multicolumn{4}{|c|}
{\rule[0cm]{0cm}{0cm}
$\sqrt s_{pp}=10$ TeV}
 \\ \hline  \hline
\rule[0cm]{0cm}{0cm}
$130$ &  $74.3$  &  $26.0$  &  $14.6(10.2)[8.3]$ \\ \hline
\rule[0cm]{0cm}{0cm}
$150$ &  $151.1$  &  $15.6$  &  $38.3(27.1)[22.1]$ \\ \hline
\rule[0cm]{0cm}{0cm}
$170$ &  $34.2$  &  $12.4$  &  $9.7(6.7)[5.5]$ \\ \hline
\multicolumn{4}{|c|}
{\rule[0cm]{0cm}{0cm}
$\sqrt s_{pp}=14$ TeV}
 \\ \hline  \hline
\multicolumn{4}{|c|}
{\rule[0cm]{0cm}{0cm}
$m_t=175$ GeV}
 \\ \hline 
\multicolumn{4}{c}
{\rule{0cm}{1cm}
{\Large Tab. IV}}  \\
\multicolumn{4}{c}
{\rule{0cm}{0cm}}
\end{tabular}
\end{center}
\end{table}

\vfill
\newpage

\begin{table}
\begin{center}
\begin{tabular}{|c|c|c|c|}
\hline
\multicolumn{4}{|c|}
{\rule[0cm]{0cm}{0cm}
$4\mu X$}
 \\ \hline  \hline
\rule[0cm]{0cm}{0cm}
$M_H$~(GeV) &  $S$  &  $B$/2 GeV &  $S/\sqrt B$ \\ \hline\hline
\rule[0cm]{0cm}{0cm}
$130$ &  $28.3$  &  $13.5$  &  $7.7(5.3)[4.3]$ \\ \hline
\rule[0cm]{0cm}{0cm}
$150$ &  $55.6$  &  $8.2$  &  $19.4(13.7)[11.2]$ \\ \hline
\rule[0cm]{0cm}{0cm}
$170$ &  $10.4$  &  $4.3$  &  $5.0(3.6)[3.0]$ \\ \hline
\multicolumn{4}{|c|}
{\rule[0cm]{0cm}{0cm}
$\sqrt s_{pp}=10$ TeV}
 \\ \hline  \hline
\rule[0cm]{0cm}{0cm}
$130$ &  $44.9$  &  $17.3$  &  $10.8(7.5)[6.1]$ \\ \hline
\rule[0cm]{0cm}{0cm}
$150$ &  $90.9$  &  $10.1$  &  $28.6(20.2)[16.5]$ \\ \hline
\rule[0cm]{0cm}{0cm}
$170$ &  $17.4$  &  $8.2$  &  $6.1(4.4)[3.6]$ \\ \hline
\multicolumn{4}{|c|}
{\rule[0cm]{0cm}{0cm}
$\sqrt s_{pp}=14$ TeV}
 \\ \hline  \hline
\multicolumn{4}{|c|}
{\rule[0cm]{0cm}{0cm}
$m_t=175$ GeV}
 \\ \hline 
\multicolumn{4}{c}
{\rule{0cm}{1cm}
{\Large Tab. V}}  \\
\multicolumn{4}{c}
{\rule{0cm}{0cm}}
\end{tabular}
\end{center}
\end{table}





\vfill
\newpage

\begin{figure}[p]
~\epsfig{file=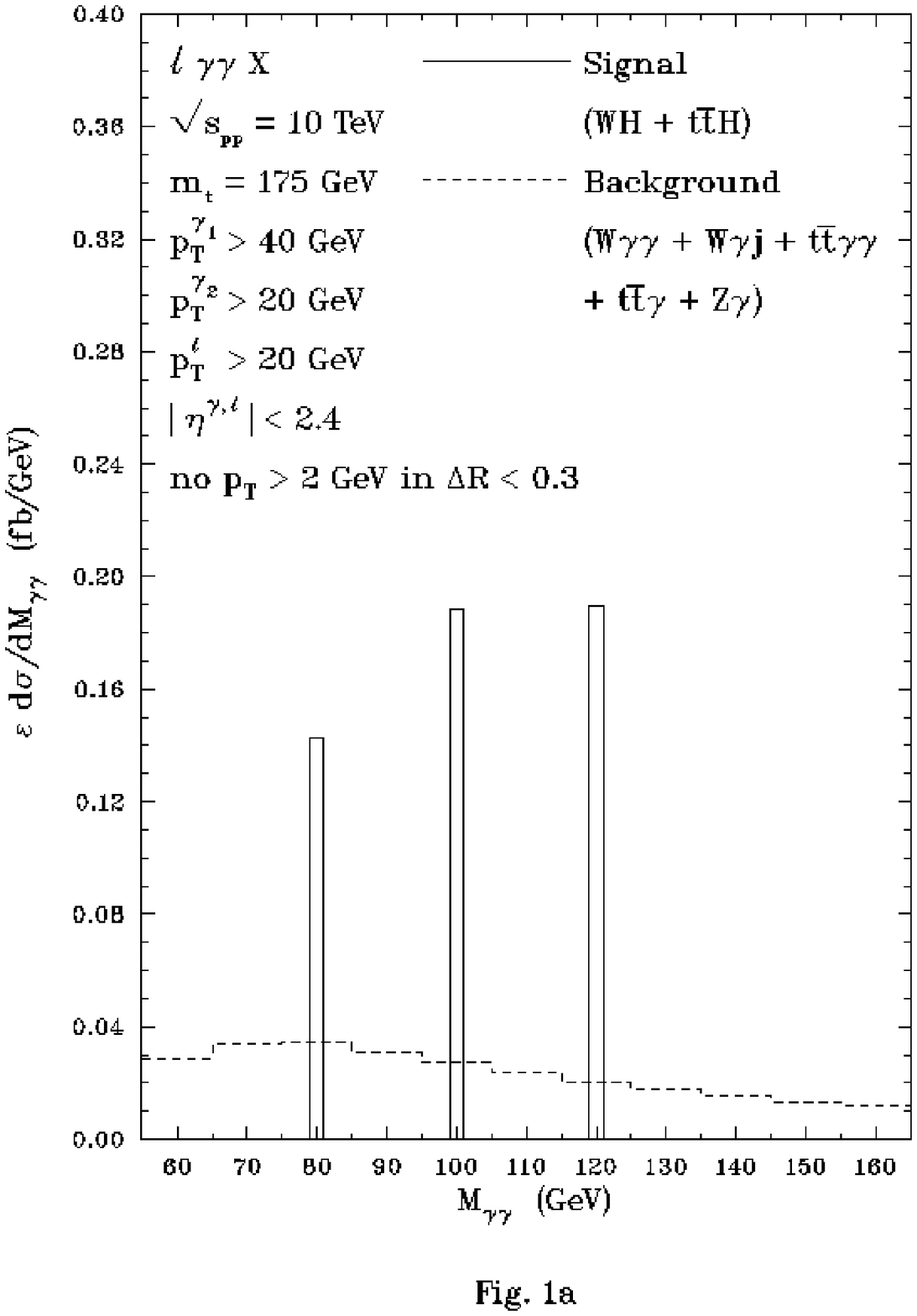,height=22cm}
\end{figure}
\stepcounter{figure}
\vfill
\clearpage

\begin{figure}[p]
~\epsfig{file=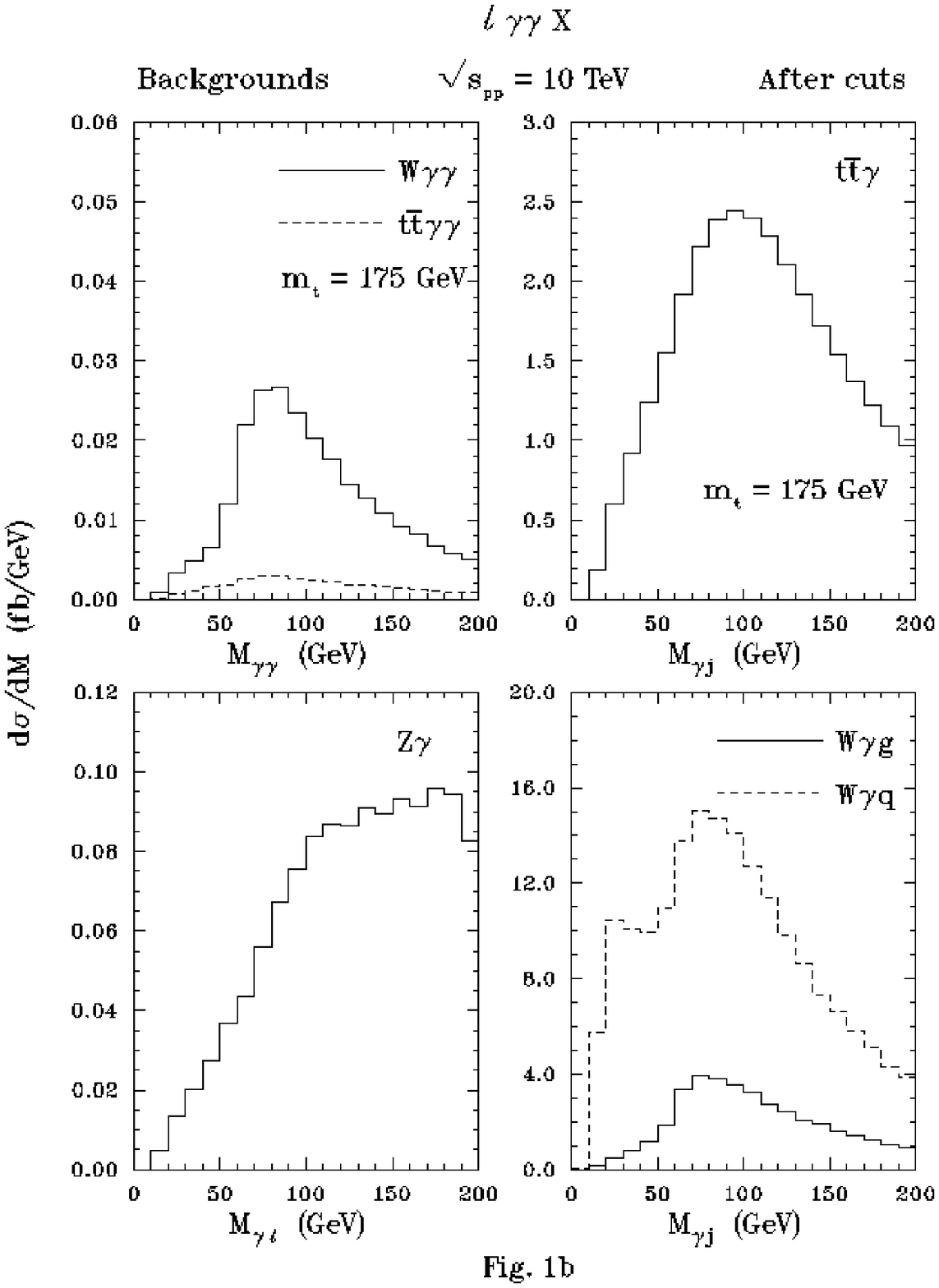,height=22cm}
\end{figure}
\stepcounter{figure}
\vfill
\clearpage

\begin{figure}[p]
~\epsfig{file=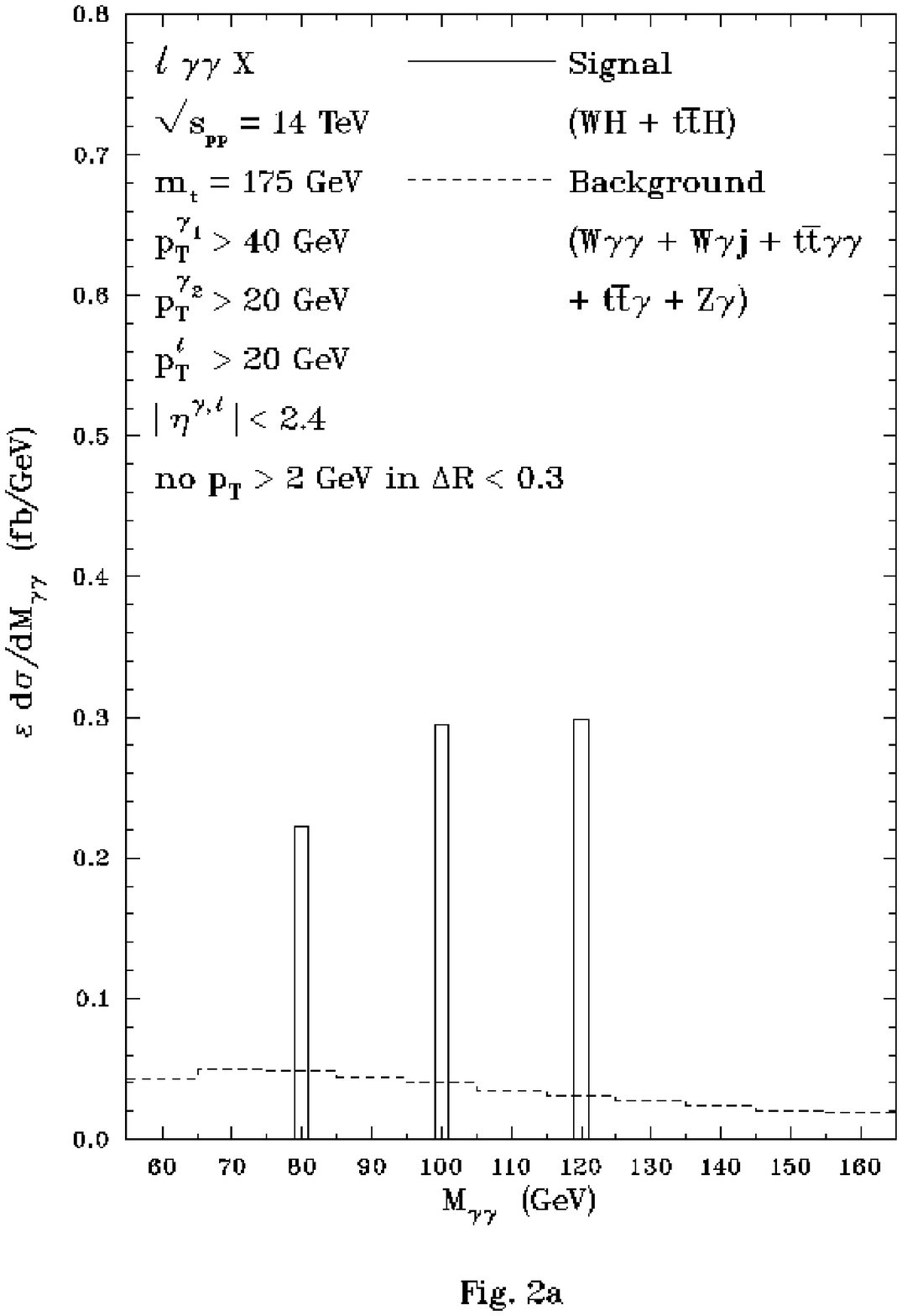,height=22cm}
\end{figure}
\stepcounter{figure}
\vfill
\clearpage

\begin{figure}[p]
~\epsfig{file=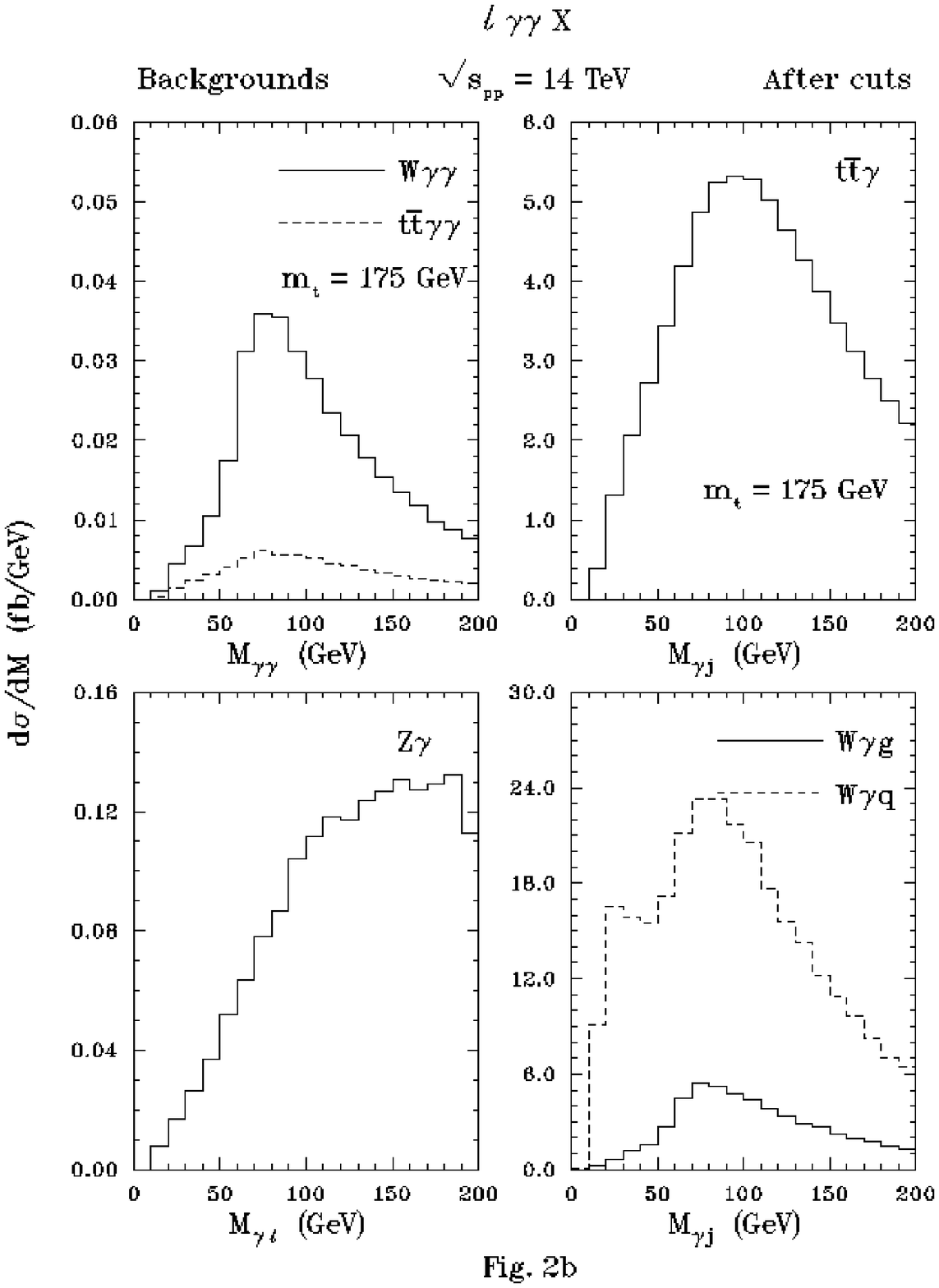,height=22cm}
\end{figure}
\stepcounter{figure}
\vfill
\clearpage

\begin{figure}[p]
~\epsfig{file=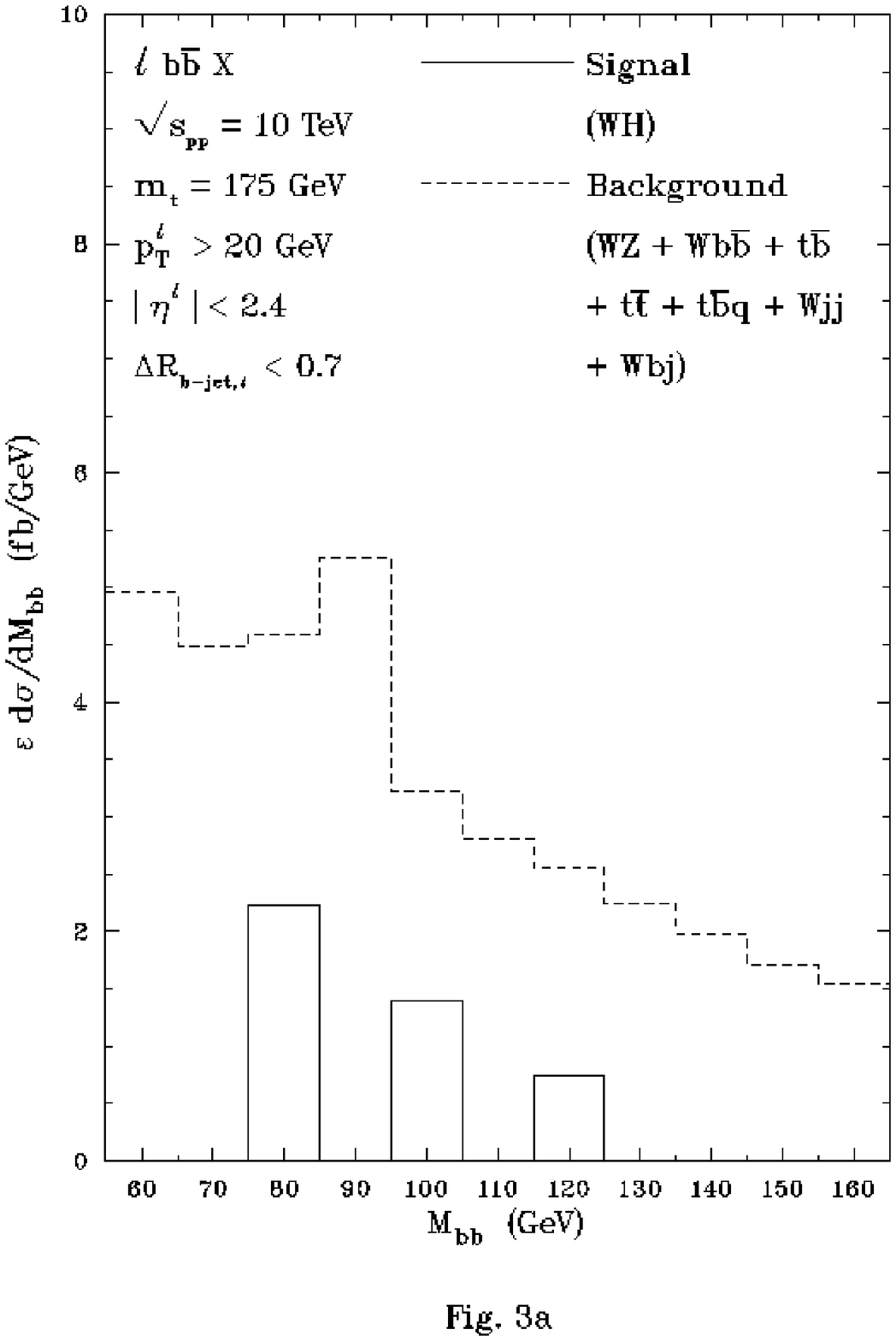,height=22cm}
\end{figure}
\stepcounter{figure}
\vfill
\clearpage

\begin{figure}[p]
~\epsfig{file=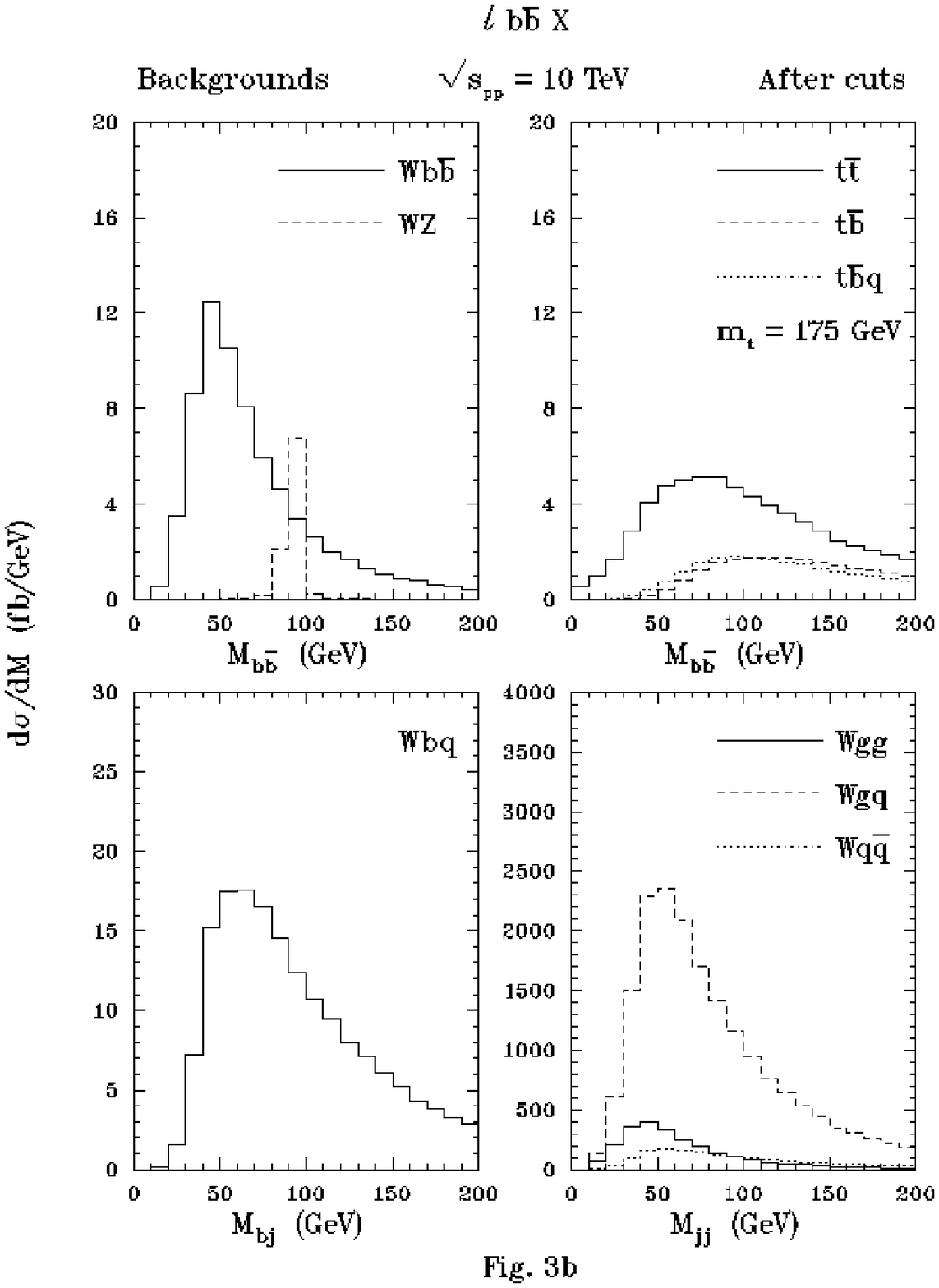,height=22cm}
\end{figure}
\stepcounter{figure}
\vfill

\begin{figure}[p]
~\epsfig{file=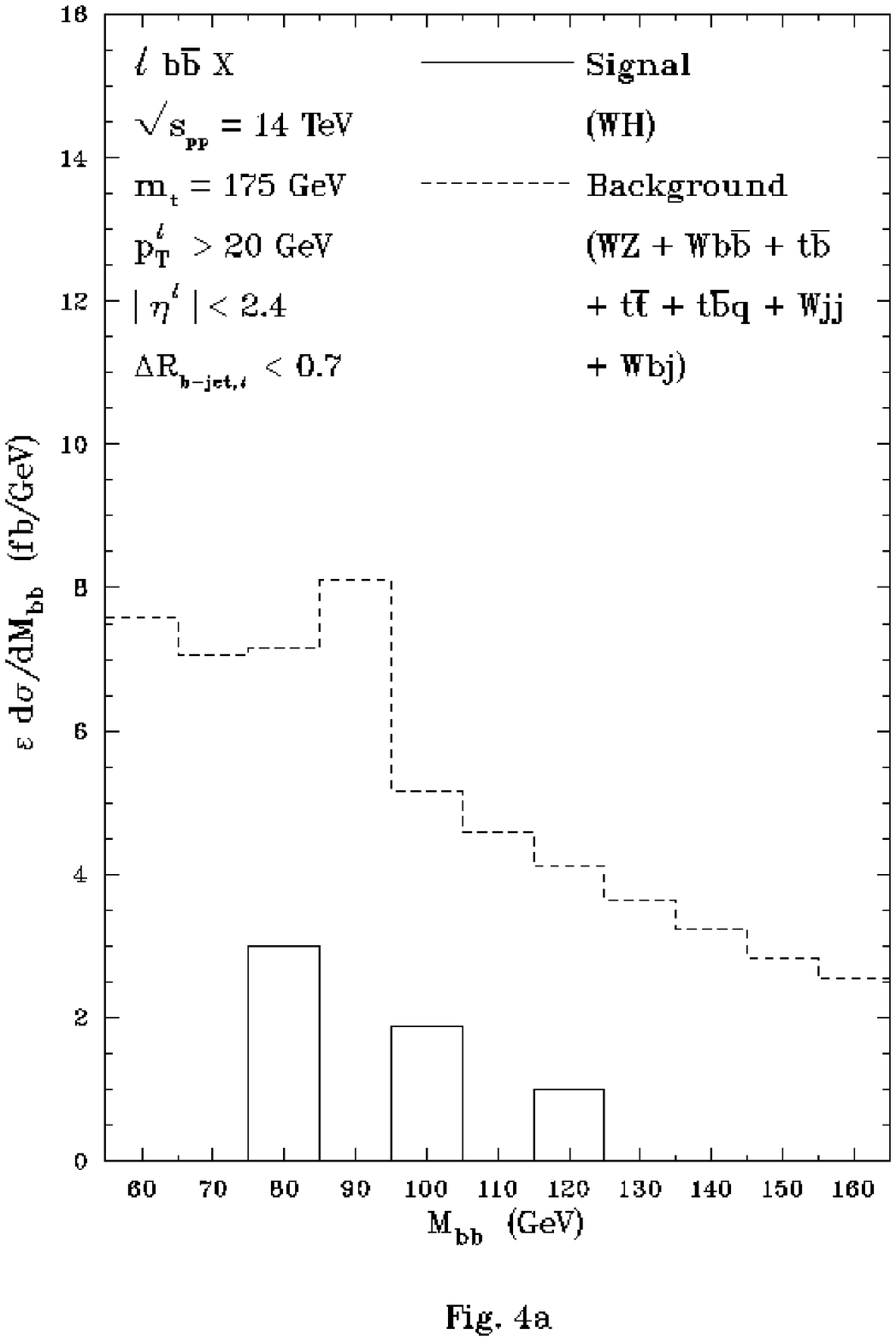,height=22cm}
\end{figure}
\stepcounter{figure}
\vfill
\clearpage

\begin{figure}[p]
~\epsfig{file=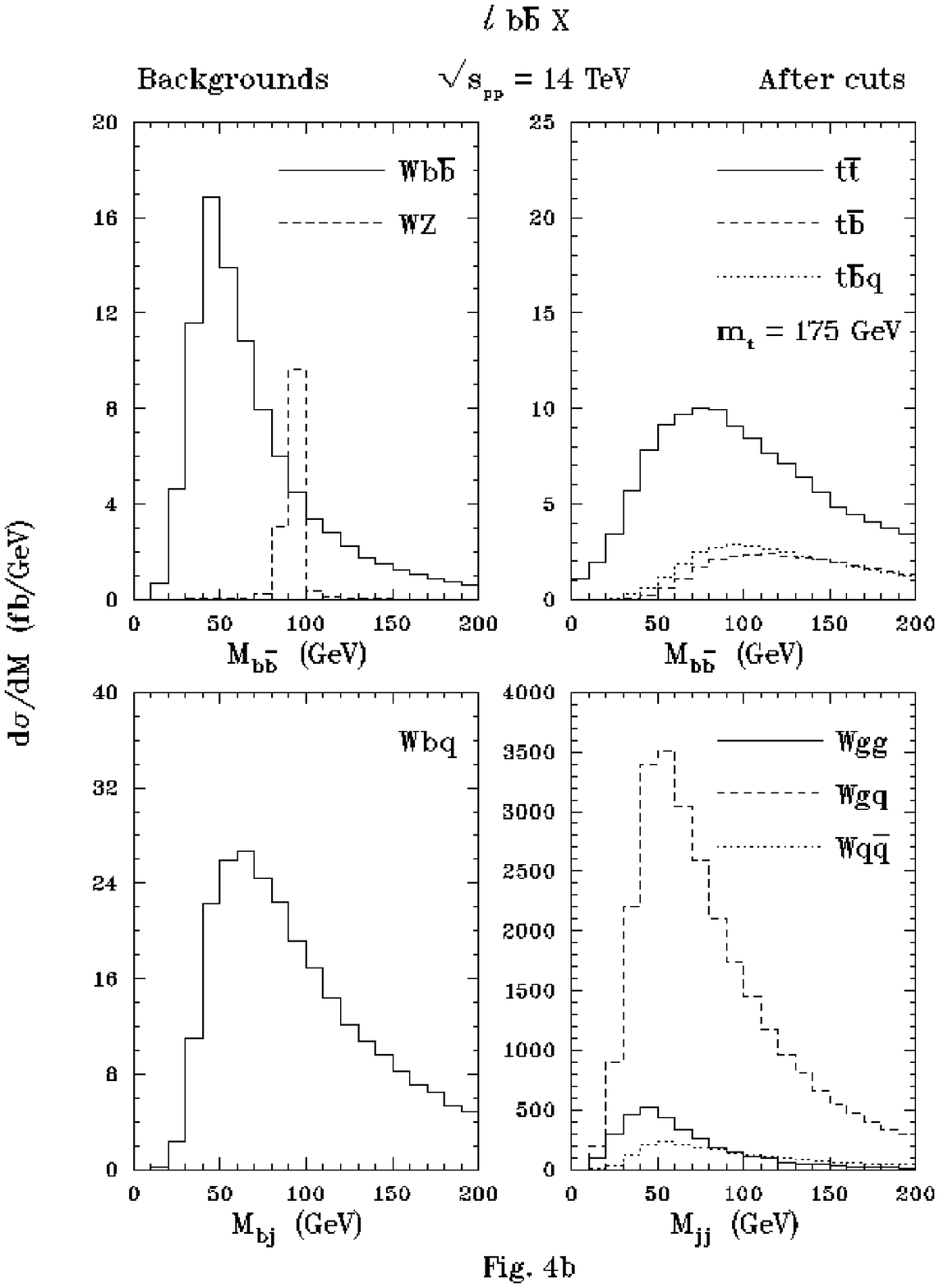,height=22cm}
\end{figure}
\stepcounter{figure}
\vfill
\clearpage

\begin{figure}[p]
~\epsfig{file=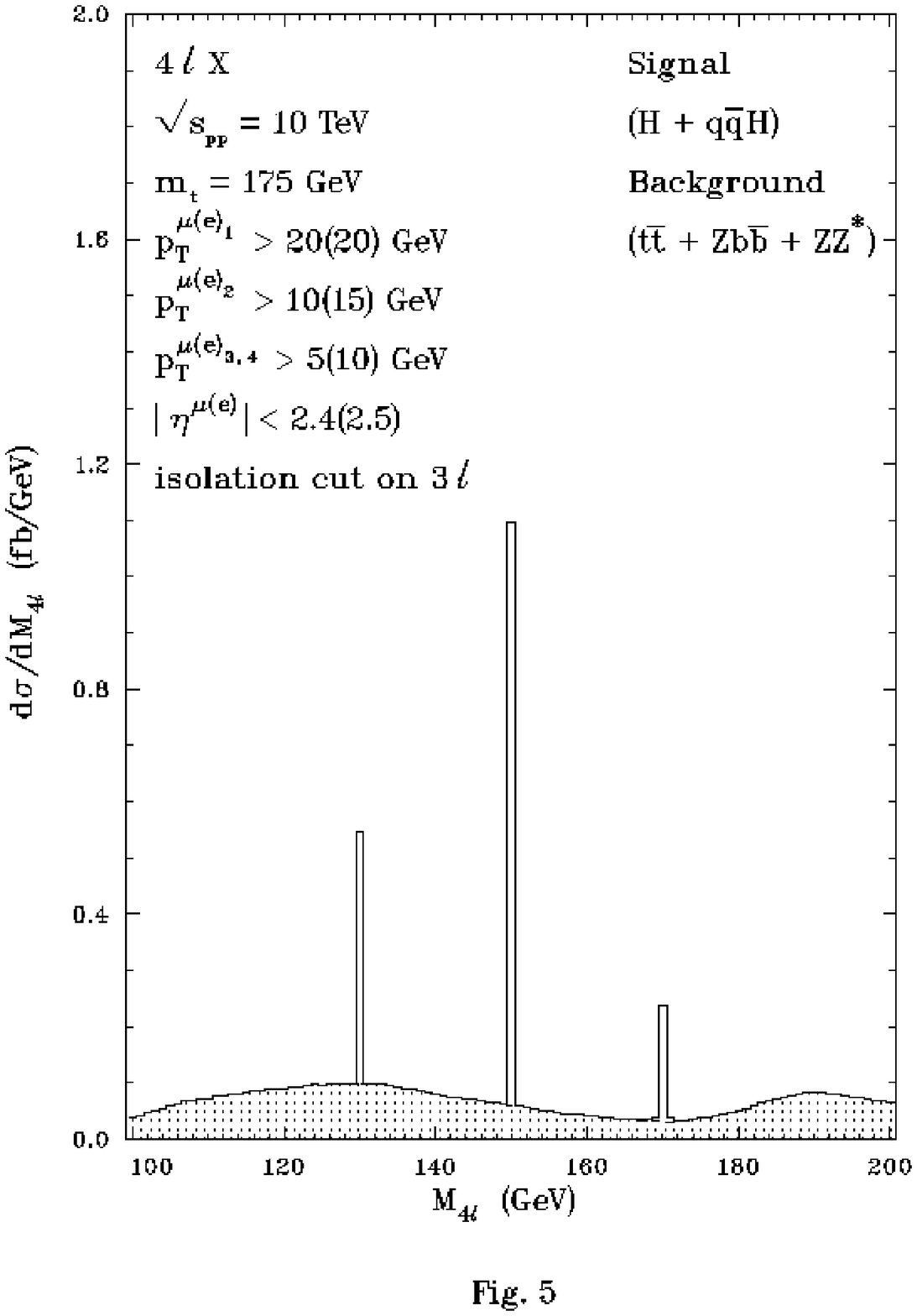,height=22cm}
\end{figure}
\stepcounter{figure}
\vfill
\clearpage

\begin{figure}[p]
~\epsfig{file=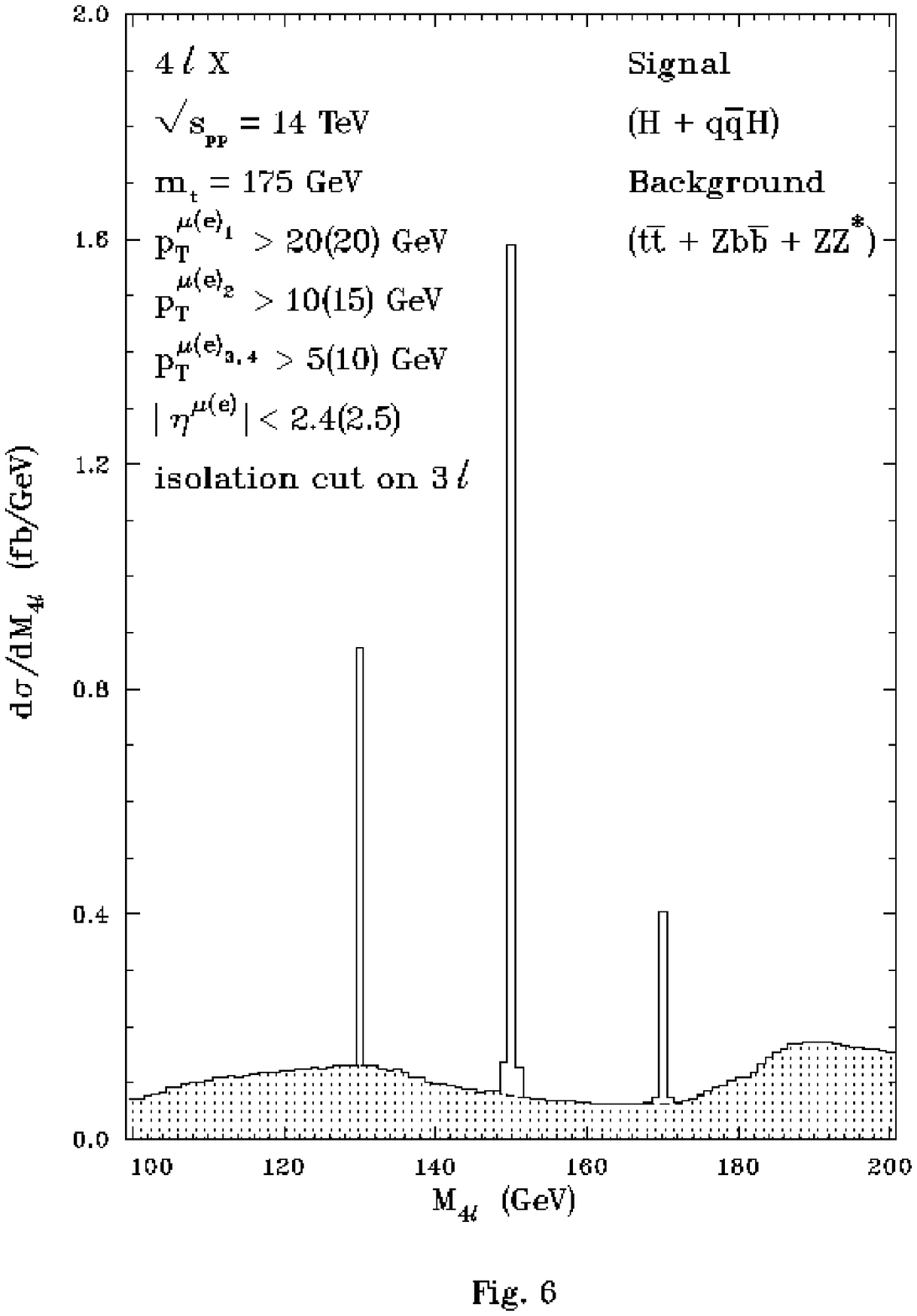,height=22cm}
\end{figure}
\stepcounter{figure}
\vfill
\clearpage

\begin{figure}[p]
~\epsfig{file=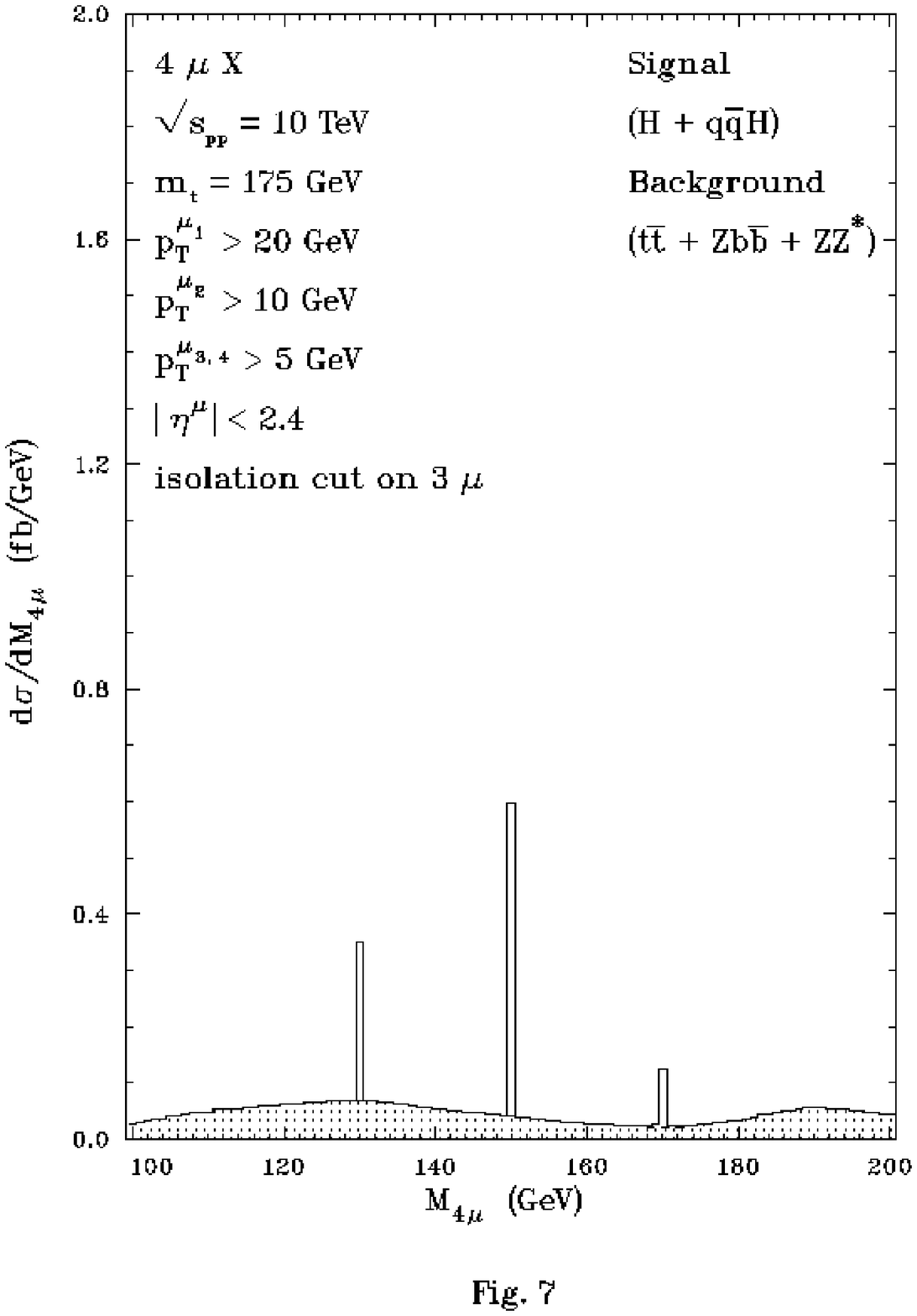,height=22cm}
\end{figure}
\stepcounter{figure}
\vfill
\clearpage

\begin{figure}[p]
~\epsfig{file=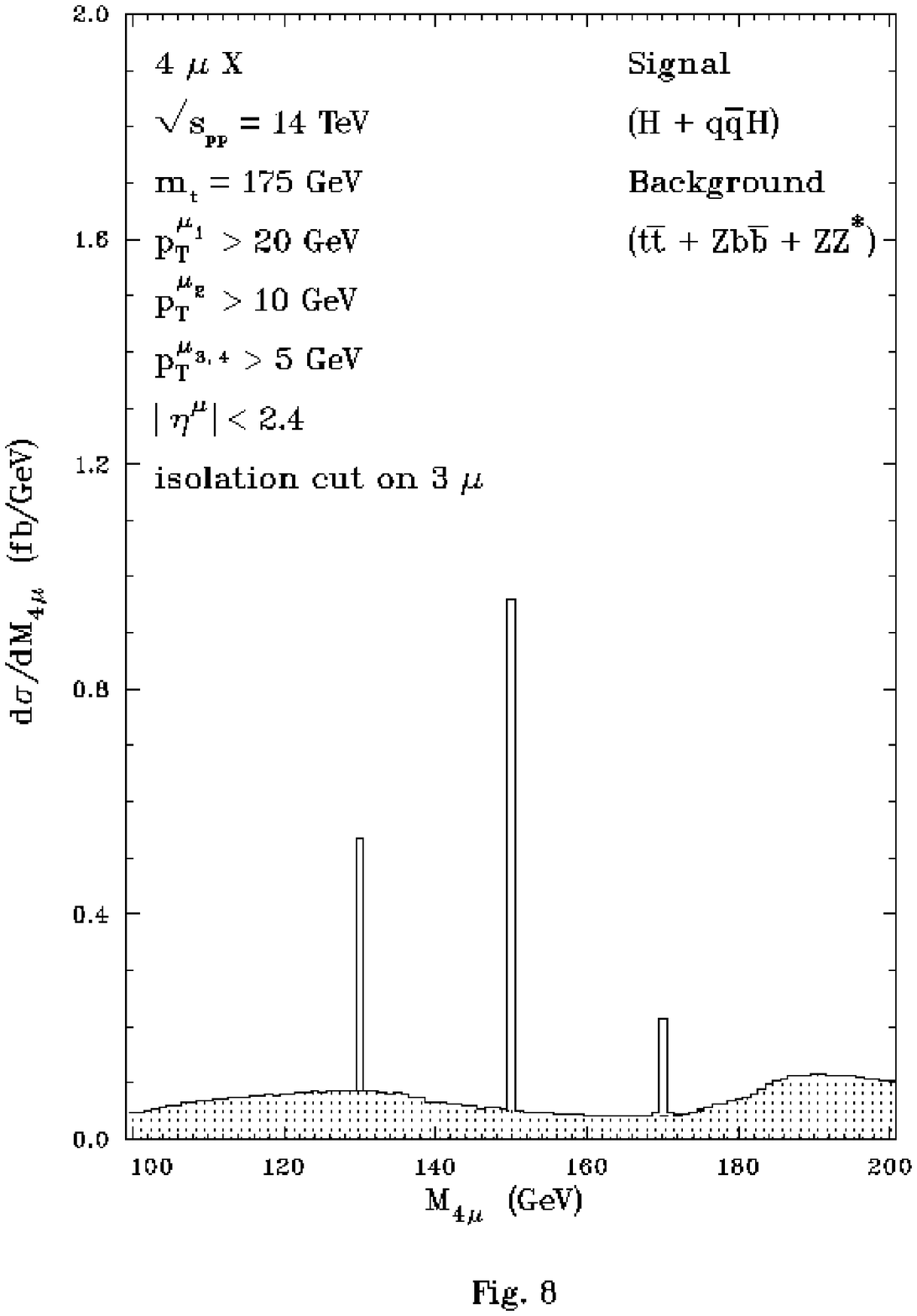,height=22cm}
\end{figure}
\stepcounter{figure}
\vfill
\clearpage

\begin{figure}[p]
~\epsfig{file=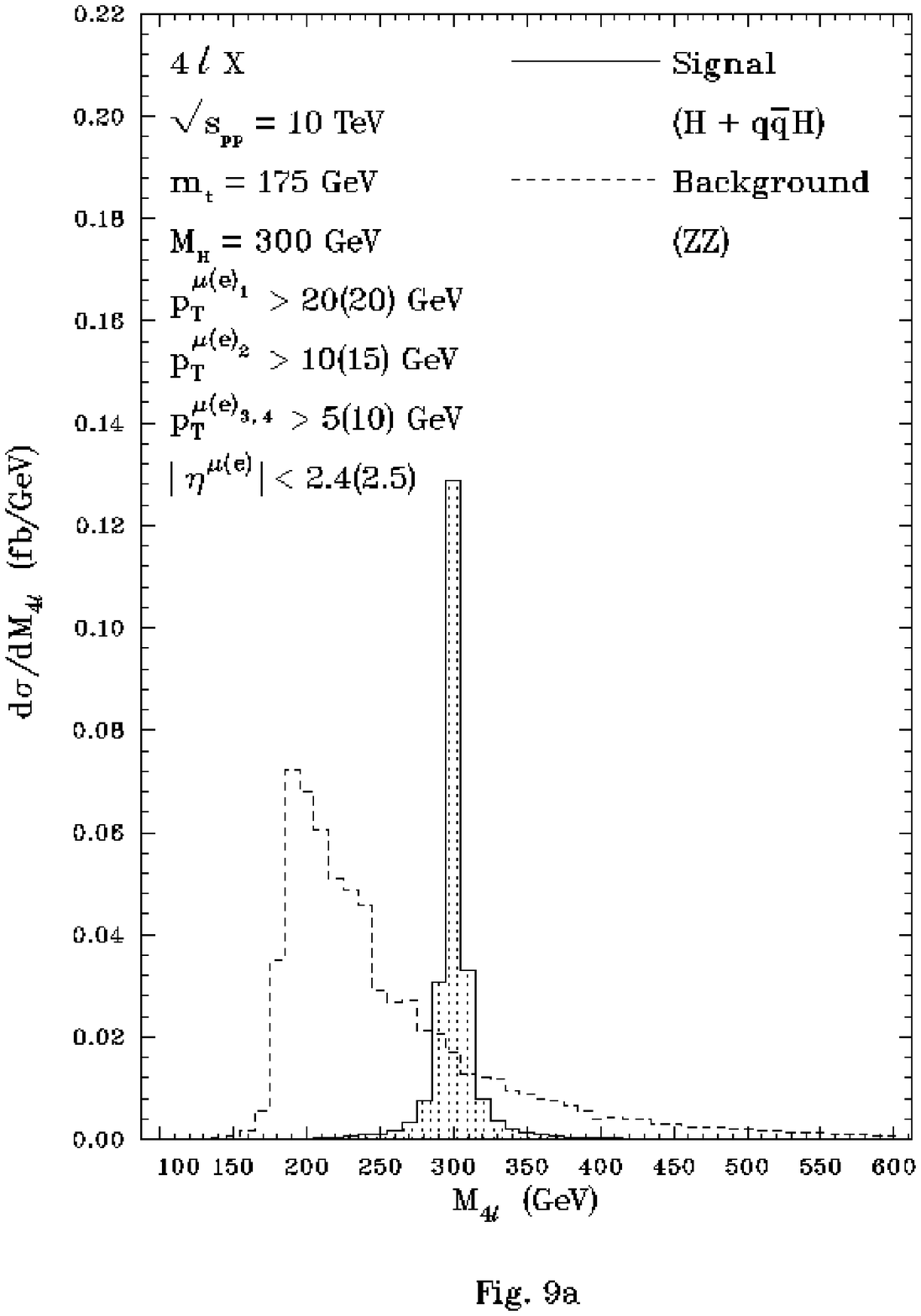,height=22cm}
\end{figure}
\stepcounter{figure}
\vfill
\clearpage

\begin{figure}[p]
~\epsfig{file=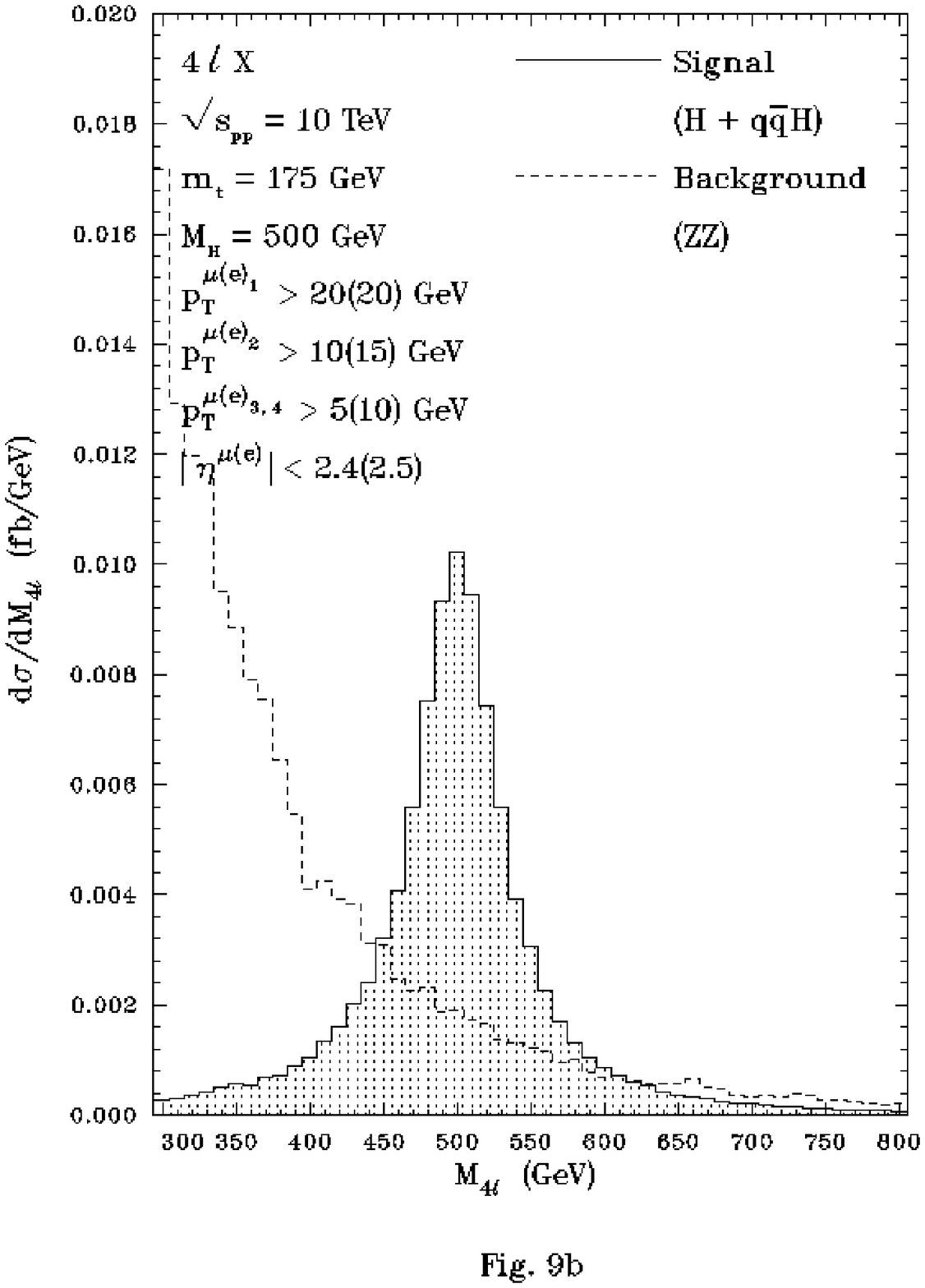,height=22cm}
\end{figure}
\stepcounter{figure}
\vfill
\clearpage

\begin{figure}[p]
~\epsfig{file=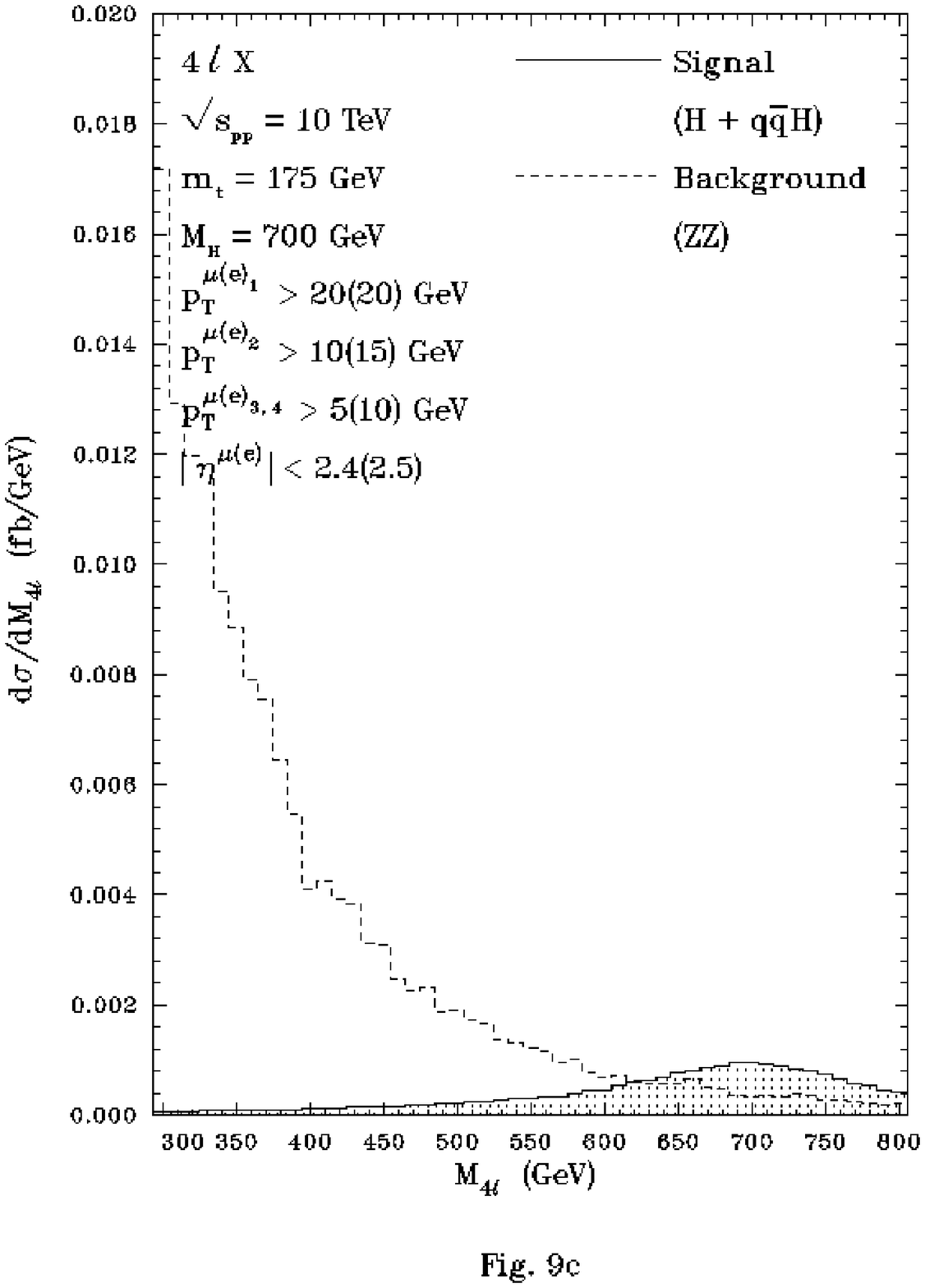,height=22cm}
\end{figure}
\stepcounter{figure}
\vfill
\clearpage

\begin{figure}[p]
~\epsfig{file=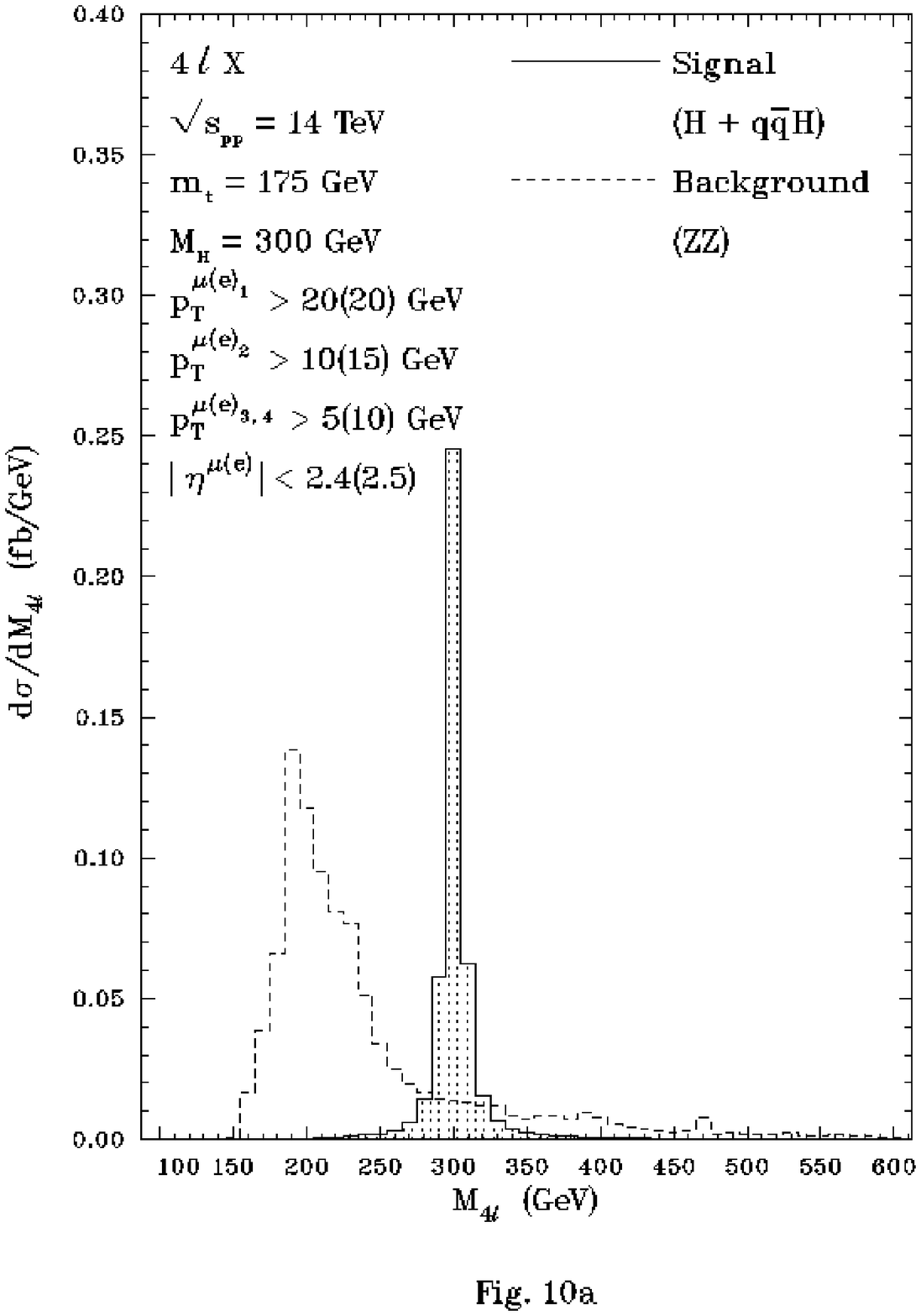,height=22cm}
\end{figure}
\stepcounter{figure}
\vfill
\clearpage

\begin{figure}[p]
~\epsfig{file=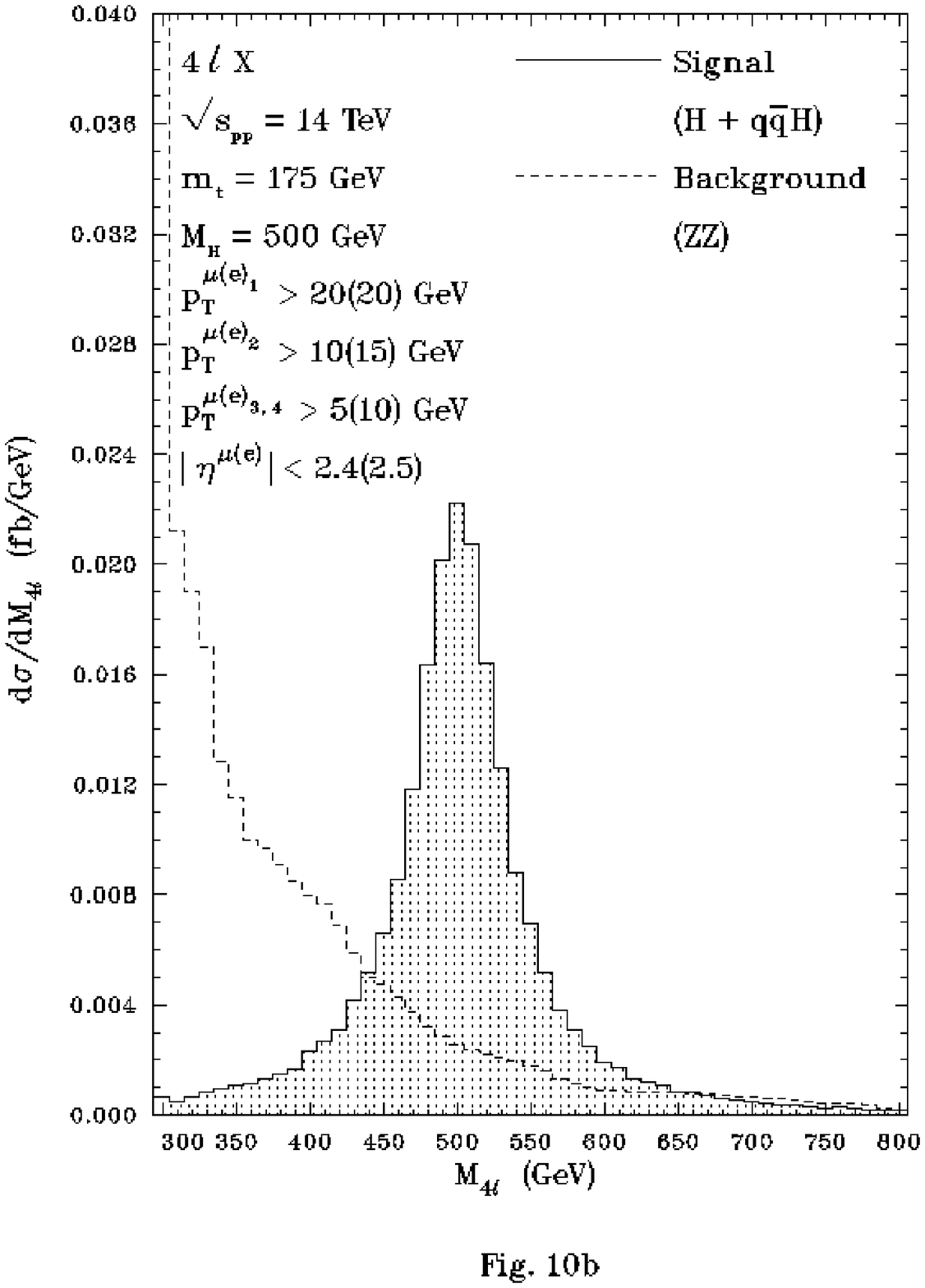,height=22cm}
\end{figure}
\stepcounter{figure}
\vfill
\clearpage

\begin{figure}[p]
~\epsfig{file=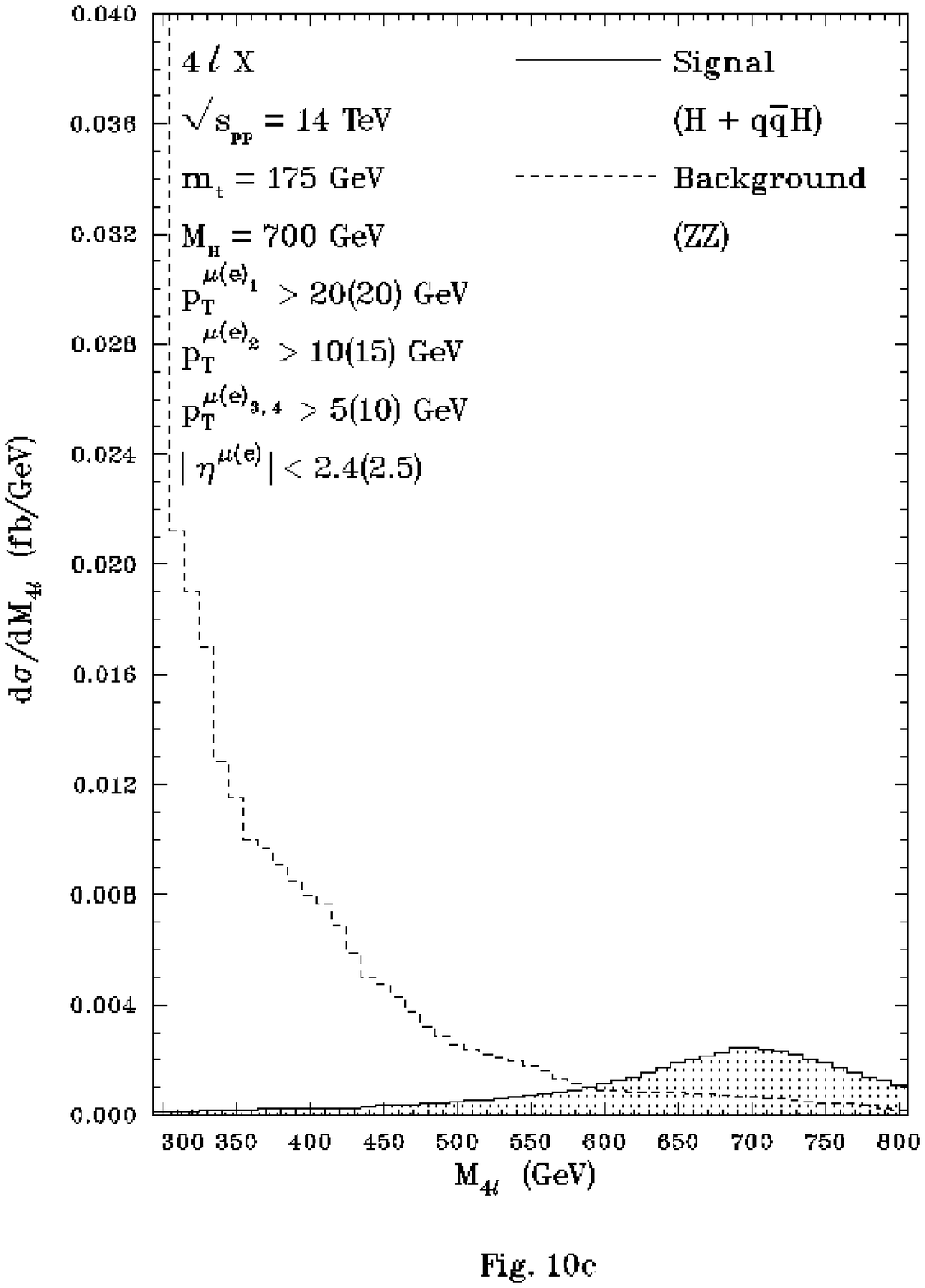,height=22cm}
\end{figure}
\stepcounter{figure}
\vfill
\clearpage

\begin{figure}[p]
~\epsfig{file=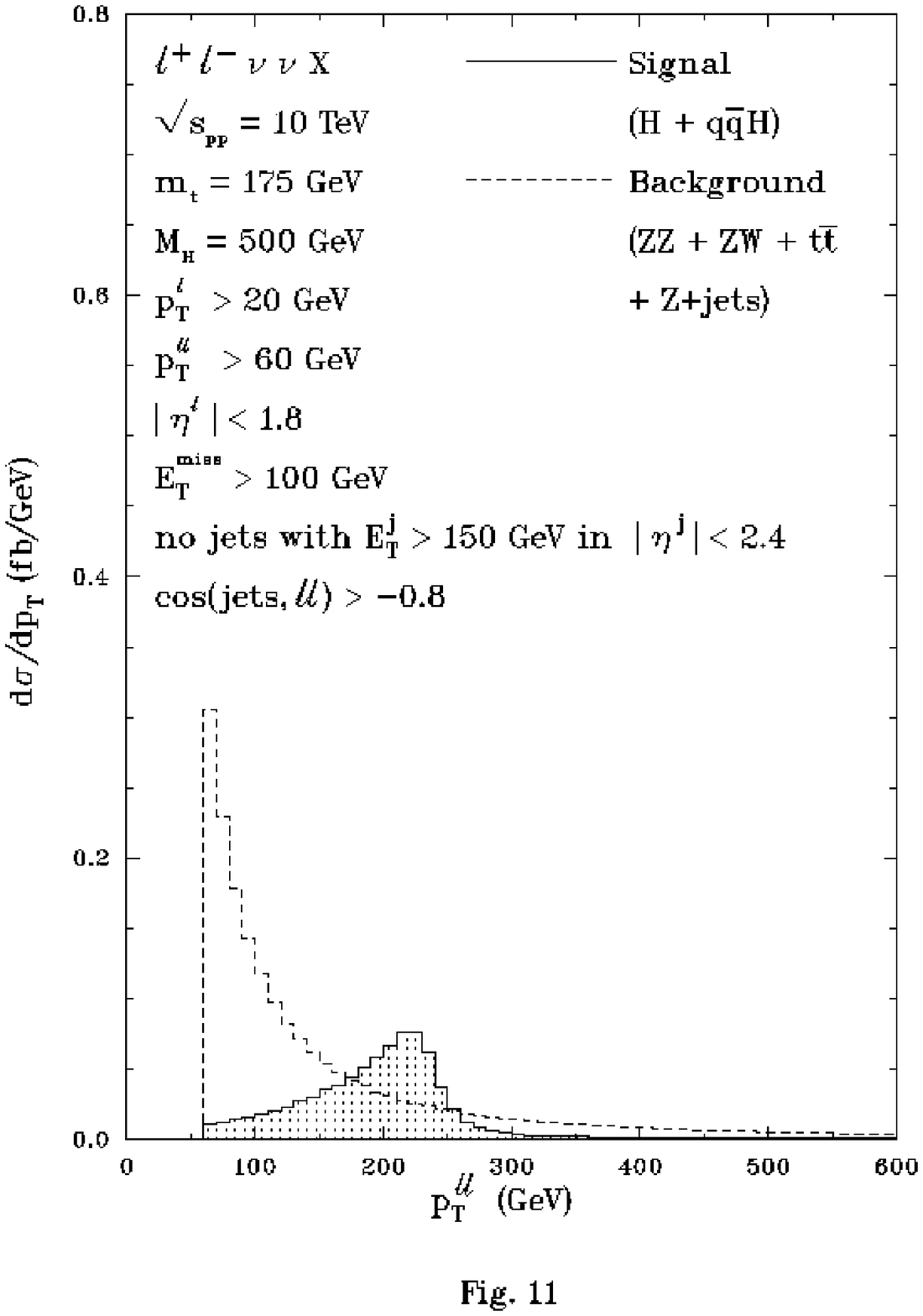,height=22cm}
\end{figure}
\stepcounter{figure}
\vfill
\clearpage

\begin{figure}[p]
~\epsfig{file=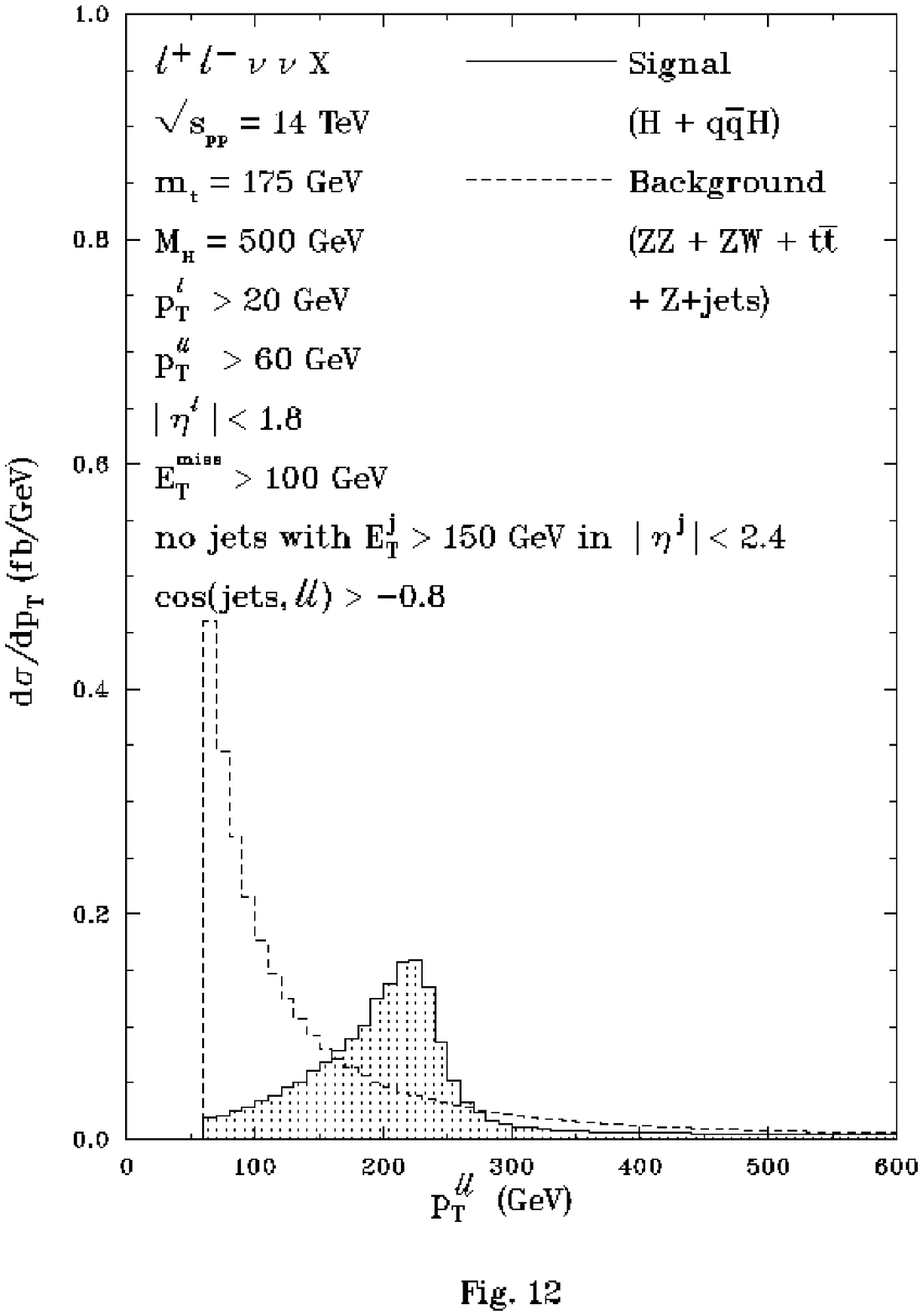,height=22cm}
\end{figure}
\stepcounter{figure}
\vfill
\clearpage


\begin{thebibliography}{99}

\bibitem{signals} Z.~Kunszt, S. Moretti and W.J. Stirling, \preprint\ 
DFTT 34/95, DTP/96/100, Cavendish-HEP-96/20, ETH-TH-96/48, November 1996.

\bibitem{DATA} See, for example:\\
Review of Particle Properties, \pr D45 1992 S1.

\bibitem{LHC} Proceedings of the ``{\it 
Large Hadron Collider Workshop}'', Aachen, 4--9 October
1990, eds. G.~Jarlskog and 
D.~Rein, Report CERN 90--10, ECFA 90--133, Geneva, 1990.  

\bibitem{ATLAS} ATLAS Technical Proposal, 
CERN/LHC/94-43 LHCC/P2 (December 1994).

\bibitem{CMS} CMS Technical Proposal, CERN/LHC/94-43 LHCC/P1 (December 1994).

\bibitem{generators} T.~Sj\"ostrand, ``High-Energy-Physics Event
Generation with PYTHIA 5.7 and jetSET 7.4'', CERN {\it preprints}
CERN-TH.7111/93 and CERN-TH.7112/93;\\
G. Marchesini, B.R. Webber {\it et al.}, {\it Comput. Phys. Commun.} 
 {\bf 67} (1992) 465;\\
GEANT, Detector Description and Simulation Tool, CERN Program Library Q123.

\bibitem{MadGraph} T.~Stelzer and W.F.~Long, {\it Comp. Phys. Comm.} {\bf 81} 
              (1994) 357.

\bibitem{helas} E.~Murayama, I.~Watanabe and K.~Hagiwara, HELAS: HELicity
                Amplitude Subroutines for Feynman Diagram Evaluations,
                {\it KEK Report} 91-11, January 1992.
  
\bibitem{Vegas} G.P.~Lepage, {\it Jour. Comp. Phys.} {\bf 27} (1978) 192.

\bibitem{MRSA}  A.D.~Martin, R.G.~Roberts and W.J.~Stirling,
                \pr D50 1994 6734.

\bibitem{CDFtop} CDF~Collaboration,
{\it Phys. Rev. Lett.} {\bf 74} (1995) 2626.

\bibitem{D0top} D0 Collaboration, {\it Phys. Rev. Lett.} {\bf 74} (1995) 2632.

\bibitem{newtop} A. Caner, Presented at ``{\it Rencontres du Physique de la
Valle d'Aoste}'', March 1996;\\
M. Narain, Presented at the `{\it Rencontres du Physique de
la Valle d'Aoste}'',
March 1996;\\
P. Grannis, talk presented at ICHEP96,
Warsaw, Poland, 25-31 July 1996, to appear in the proceedings.

\bibitem{widthtopSM} J.H.~K\"uhn, {\it Act. Phys. Pol.} {\bf B12} (1981) 347;\\
                     J.H.~K\"uhn, {\it Act. Phys. Austr. Suppl. } {\bf XXIV}
                     (1982) 203.

\bibitem{lower} J.P. Martin, talk
presented at ICHEP96, Warsaw, Poland, 
25-31 July 1996, to appear in the proceedings.

\bibitem{blondel} See, for example:\\
A. Blondel, plenary talk
presented at ICHEP96, Warsaw, Poland,
25-31 July 1996, to appear in the proceedings.

\bibitem{dittdrei} M.~Dittmar and H.~Dreiner, \preprint\ RAL-96-049
August 1996.

\bibitem{last} D.~Froidevaux and E.~Richter-Was, 
\zp C67 1995 213.

\bibitem{ATL17} L.~Fayard and G.~Unal, EAGLE Internal Note PHYS-NO-001
and Addenda 1 \& 2.

\bibitem{gamgam} C.~Seez  {\it et al.}, in Ref.~\cite{LHC}.

\bibitem{GattoeVolpe} A.~Ballestrero and E.~Maina, {\it Phys. Lett.} 
{\bf B268} (1992) 437.

\bibitem{primo} E.~Maina and S.~Moretti, \pl B286 1992 370.

\bibitem{Wjj} V.~Barger, T.~Han, J.~Ohnemus and D.~Zeppenfeld,
{\it Phys. Rev. Lett.} {\bf 62} (1989) 1971; {\it Phys. Rev.} {\bf D40} (1989) 
2888;~{\it ibidem}~{\bf D41}~(1990)~1715~(Erratum).
 
\bibitem{btagg} J.~Dai, J.F.~Gunion and R.~Vega, \prl 71 1993 2699.

\bibitem{SDC} {\it Solenoidal Detector Collaboration Technical Design Report},
              E.L.~Berger {\it et al.}, {\it Report} SDC--92--201, SSCL--SR--12

\bibitem{Tevadetect1} A.~Stange, W.~Marciano and S.S.D.~Willenbrock, 
{\it Phys. Rev.} {\bf D49} (1994) 1354; {\it ibidem} {\bf D50} (1994) 4491.

\bibitem{Tevadetect2} J.F.~Gunion and T.~Han, \pr D51 1995 1051.

\bibitem{gny} S.L.~Glashow, D.V.~Nanopoulos and A.~Yildiz, {\it Phys. Rev.}
              {\bf D18} (1978) 1724.

\bibitem{DellaNegra} M.~Della~Negra  {\it et al.}, in Ref.~\cite{LHC}.

\bibitem{CMS14} M.~Dzelalija and  R.~Kinnunen, CMS {\it internal note} CMS
TN/94-214 (1994).

\bibitem{Nigel} U.~Baur and E.W.N.~Glover, {\it Nucl. Phys.} {\bf
                B347} (1990) 12;\\
                U.~Baur and E.W.N.~Glover, in Ref.~\cite{LHC}.

\bibitem{ggZbb} R.~Kleiss and  J.J.~van~der~Bij, in Ref.~\cite{LHC}. 

\bibitem{ggZZ} E.W.N.~Glover and J.J.~van~der~Bij, \np B321 1989 561.

\bibitem{CMS2223}  N.~Stepanov and A.~Starodumov, CMS {\it internal note} CMS
                   TN/92-49 (1992);\\ 
                   N.~Stepanov, CMS {\it internal note} CMS
                   TN/93-87 (1993).

\bibitem{ATLAS3839} I.~Zuckerman {\it et al.}, ATLAS  {\it internal note} 
                    PHYS-NO-007 (1992);\\
                    M.~Bosman and M.~Nessi, ATLAS  {\it internal note} 
                    PHYS-NO-050 (1994).

\bibitem{breaks} A.~Hasenfratz, K.~Jansen, C.~Lang, T.~Neuhaus and H.~Yoneyama,
\pl B199 1987 531; \\
J.~Kuti, L.~Liu and Y. Shen, \prl  61 1988 678; \\
M. L\"uscher and P. Weisz, \np B318 1989 705.

\bibitem{KfacVH} T.~Han and S.S.D.~Willenbrock, \pl B273 1991 167.

\bibitem{ioeben} B.K. Bullock and S. Moretti, in preparation.

\end{thebibliography}
\end{document}